\documentclass[prd,superscriptaddress,amsfonts,amssymb,amsmath,showpacs,twocolumn]{revtex4-2}
\usepackage{bm}
\usepackage{amsfonts}
\usepackage{latexsym}
\usepackage[latin1]{inputenc}
\usepackage{graphicx}
\usepackage{amsmath}
\usepackage{palatino}
\usepackage{mathpazo}
\usepackage{textcomp}
\usepackage{float}
\usepackage{booktabs}
\usepackage{dcolumn}
\usepackage{multirow}
\usepackage{ragged2e}
\usepackage{hyperref}
\hypersetup{colorlinks,citecolor=blue}
\usepackage{amsmath}
\usepackage{xcolor}
\usepackage{orcidlink}
\usepackage[caption=false]{subfig}
\usepackage{commath}
\def\jnl@style{\it}
\def\aaref@jnl#1{{\jnl@style#1}}

\def\aaref@jnl#1{{\jnl@style#1}}

\def\aj{\aaref@jnl{AJ}}                   % Astronomical Journal
\def\apj{\aaref@jnl{ApJ}}                 % Astrophysical Journal
\def\apjl{\aaref@jnl{ApJ}}                % Astrophysical Journal, Letters
\def\apjs{\aaref@jnl{ApJS}}               % Astrophysical Journal, Supplement
\def\apss{\aaref@jnl{Ap\&SS}}             % Astrophysics and Space Science
\def\aap{\aaref@jnl{A\&A}}                % Astronomy and Astrophysics
\def\aapr{\aaref@jnl{A\&A~Rev.}}          % Astronomy and Astrophysics Reviews
\def\aaps{\aaref@jnl{A\&AS}}              % Astronomy and Astrophysics, Supplement
\def\mnras{\aaref@jnl{Mon.~Not.~Roy.~Astron.~Soc.}}             % Monthly Notices of the RAS
\def\prd{\aaref@jnl{Phys.~Rev.~D}}        % Physical Review D
\def\prc{\aaref@jnl{Phys.~Rev.~C}}  % Physical Review C
\def\prl{\aaref@jnl{Phys.~Rev.~Lett.}}    % Physical Review Letters
\def\qjras{\aaref@jnl{QJRAS}}             % Quarterly Journal of the RAS
\def\skytel{\aaref@jnl{S\&T}}             % Sky and Telescope
\def\ssr{\aaref@jnl{Space~Sci.~Rev.}}     % Space Science Reviews
\def\zap{\aaref@jnl{ZAp}}                 % Zeitschrift fuer Astrophysik
\def\nat{\aaref@jnl{Nature}}              % Nature
\def\aplett{\aaref@jnl{Astrophys.~Lett.}} % Astrophysics Letters
\def\apspr{\aaref@jnl{Astrophys.~Space~Phys.~Res.}} % Astrophysics Space Physics Research
\def\physrep{\aaref@jnl{Phys.~Rep.}}      % Physics Reports
\def\physscr{\aaref@jnl{Phys.~Scr}}       % Physica Scripta
\def\commat{\aaref@jnl{Comm.~Math.~Phys.}}              % Communications in Mathematical Physics
\def\science{\aaref@jnl{Science}}               % Science
\def\cqg{\aaref@jnl{Classical Quant.~Grav.}}            % Classical and Quantum Gravity
\def\jpcs{\aaref@jnl{JPCS}}                                     % Journal of Physics Conference Series
\def\ijmpd{\aaref@jnl{Int.~J.~Mod.~Phys.~D}}                    % International Journal of Modern Physics D
\def\grg{\aaref@jnl{Gen.~Relat.~Gravit.}}               % General Relativity and Gravitation
\def\rpp{\aaref@jnl{Rep.~Prog.~Phys.}}          % Reports on Progress in Physics
\def\npa{\aaref@jnl{Nucl.~Phys.~A}}        % Nuclear Physics A
\def\lrr{\aaref@jnl{Living Rev.~Rel.}}                   % Living reviews in relativity
\def\jcap{\aaref@jnl{J.~Cosmology Astropart.~Phys.}}    % Journal of cosmology and astroparticle physics
\def\rmp{\aaref@jnl{Rev.~Mod.~Phys.}}   %Reviews of modern physics
\def\epjc{\aaref@jnl{Eur.~Phys.~J.~C}}

\allowdisplaybreaks[1]
\begin{document}

\color{black}       %% For one column

\title{Viscous fluid cosmology in symmetric teleparallel gravity}

\author{Raja Solanki\orcidlink{0000-0001-8849-7688}}
\email{rajasolanki8268@gmail.com}
\affiliation{Department of Mathematics, Birla Institute of Technology and
Science-Pilani,\\ Hyderabad Campus, Hyderabad-500078, India.}
\author{Dheeraj Singh Rana\orcidlink{0000-0002-4401-8814}}
\email{drjrana2@gmail.com}
\affiliation{Department of Mathematics, Birla Institute of Technology and
Science-Pilani,\\ Hyderabad Campus, Hyderabad-500078, India.}
\author{Sanjay Mandal\orcidlink{0000-0003-2570-2335}}
\email{sanjaymandal960@gmail.com}
\affiliation{Department of Mathematics, Birla Institute of Technology and
Science-Pilani,\\ Hyderabad Campus, Hyderabad-500078, India.}
\author{P.K. Sahoo\orcidlink{0000-0003-2130-8832}}
\email{pksahoo@hyderabad.bits-pilani.ac.in}
\affiliation{Department of Mathematics, Birla Institute of Technology and
Science-Pilani,\\ Hyderabad Campus, Hyderabad-500078, India.}

%%%%%%%%%%%%%%%%%%%%%%%%%%%%%%%%%%%%  DATE  %%%%%%%%%%%%%%%%%%%%%%%%%%%%%%%%%%%%

\date{\today}
\begin{abstract}
In this manuscript, we analyze the viscous fluid cosmological model in the framework of recently proposed $f(Q)$ gravity by assuming three different forms of bulk viscosity coefficients, specifically, $(i)\zeta =\zeta_{0}+\zeta_{1}\left( \frac{\dot{a}}{a}\right) +\zeta_{2}\left( \frac{{\ddot{a}}}{\dot{a}}\right) $, $(ii)\zeta =\zeta_{0}+\zeta_{1}\left( \frac{\dot{a}}{a}\right)$, and $(iii)\zeta =\zeta_{0}$ and a linear $f(Q)$ model, particularly, $f(Q)=\alpha Q$ where $\alpha \neq 0$ is free model parameter. We estimate the bulk viscosity coefficients and the model parameter values using the combined H(z)+Pantheon+BAO data set. We study the asymptotic behavior of our cosmological bulk viscous model by utilizing the phase space method. We find that corresponding to all three cases, our model depicts the evolution of the universe from matter dominated decelerated epoch (a past attractor) to a stable de-sitter accelerated epoch (a future attractor). Further, we study the physical behavior of effective pressure, effective equation of state (EoS), and the statefinder parameters. We find that the pressure component in the presence of bulk viscosity shows negative behavior and the effective EoS parameter predicts the accelerated expansion phase of the universe for all three cases. Moreover, we obtain that the trajectories of our model lie in the quintessence region and it converges to the $\Lambda$CDM fixed point in the far future. We find that the accelerated deSitter like phase comes purely from the $\bar{\zeta}_0$ case without any geometrical modification to GR. Moreover, we find that the late-time behavior of all three cases of viscosity coefficients are identical. Further, we consider a non-linear $f(Q)$ model, specifically, $f(Q)=-Q+\beta Q^2$ and then we analyze the behavior of model using dynamical approach. We find that the late-time behavior of the considered non-linear model $f(Q)=-Q+\beta Q^2$ with $\beta \leq 0$ is similar to the linear case, whereas for the case $\beta > 0$ results are quite different. 
\end{abstract}

\maketitle


\section{Introduction}\label{sec1}
\justify
The discovery of the accelerating behavior of the cosmos is one of the challenging issues. Several cosmic observations such as Supernovae searches \cite{Riess,Perlmutter}, BAO \cite{D.J.,W.J.}, WMAP \cite{C.L.,D.N.}, and CMBR \cite{R.R.,Z.Y.} are the evidence for this accelerating behavior. The best fit model corresponding to these observational aspects is the $\Lambda$CDM model, where $\Lambda$ is the dark energy component responsible for the aforementioned acceleration. Despite its observational compatibility, the $\Lambda$CDM model has some well known delicate issues, namely cosmic coincidence and cosmological constant problem \cite{COP}. A way to bypass the above mentioned dark energy issue is to modify the geometry of the standard general relativity (GR) by proposing a more generic action to describe gravitational interactions. This approach can be seen within the references \cite{L.A.,SA,R.F.} and is commonly known as modified theories of gravity.

In this work, we are going to present our analysis and outcomes under the framework of recently proposed $f(Q)$ gravity \cite{J.B.}. The standard GR corresponds to the spacetime having non-zero curvature tensor while vanishing non-metricity and torsion, the teleparallel equivalent of GR (TEGR) corresponds to the spacetime having non-zero torsion tensor while vanishing non-metricity and curvature, and the symmetric teleparallel equivalent of GR (STEGR) corresponds to the spacetime having non-zero non-metricity tensor while vanishing torsion and curvature. The $f(R)$, $f(T)$, and $f(Q)$ gravity  are straightforward generalizations of GR, TEGR, and STEGR respectively \cite{GTG}. One notable advantage of $f(Q)$ gravity is that its field equations are of second order in scale factor so that one can assume the Lagrangian of higher order without having any issues. Another advantage of $f(Q)$ gravity is that it holds Bianchi's identity automatically, while in the case of $f(T)$ gravity, the presence of the anti-symmetric part becomes hindered. Furthermore, some generalizations of $f(Q)$ theory have been proposed in the literature which is commonly known as $f(Q,T)$ theory \cite{m1} and Weyl type $f(Q,T)$ theory \cite{m2}. Recently, several interesting cosmological and astrophysical implications of $f(Q)$ gravity has been appeared, for instance, Cosmography \cite{SM1}, Energy conditions \cite{SM2}, Gravitational waves \cite{e3,e4,n1}, Black hole solution \cite{e7},  Quantum cosmology \cite{ND}, General covariant symmetric teleparallel cosmology \cite{e9}, Wormhole solution \cite{e8}, Evidence that non-metricity $f(Q)$ gravity can challenge $\Lambda$CDM \cite{e1}, Bouncing Cosmology \cite{bb1}, Hamiltonian analysis and ADM formulation \cite{hh1}, Statefinder Analysis of $f(Q)$ cosmology \cite{rs1}, and the Cosmological perturbations \cite{jimenez/2020}.

\justify Earlier, to study the early inflationary phase of cosmic evolution without invoking any dark energy component, viscosity in the cosmic fluid has been considered. Eckart proposed the non-casual theory of viscosity by assuming the first order deviations from the equilibrium \cite{C.E.}. Later, Israel and Stewart proposed the casual theory of viscosity by assuming the second order deviations from the equilibrium \cite{W.I.,W.I.-2,W.I.-3}. To investigate the late time cosmic scenario, the casual theory has been used. When we consider the deviations of the second order, there are two different coefficients of viscosity that comes into the picture, namely shear and bulk viscosity. The velocity gradient related to the shear viscosity vanishes in the homogeneous universe. Therefore in the case of a homogeneous and isotropic FLRW universe, only the bulk viscosity coefficient comes into play. When the cosmic fluid expands with the expansion of the universe, then the measure of pressure required to recover the thermal stability can be regarded as the bulk viscosity. Basically, the geometrical modification in the Einstein-Hilbert action of GR justifies the cosmic expansion, whereas the viscosity coefficients contribute to the pressure term to drive the cosmic acceleration. Recently, the effect of bulk viscosity in cosmic evolution has been intensively investigated, for instance, one can check the references \cite{IB-1,IB-2,IB-3,IB-4,IB-5,JM,AVS,MAT}.

In this work, we are going to analyze the role of different bulk viscosity coefficients in cosmic evolution by considering $f(Q)$ gravity. The paper is organized as follows. In Sec \ref{sec2}, we present the geometrical aspects of non-metricity $f(Q)$ gravity. In Sec \ref{sec3}, we present the flat FLRW Universe in symmetric teleparallel cosmology with bulk viscous matter. In Sec \ref{sec4}, we derive the expressions for the Hubble parameter by assuming three different bulk viscosity coefficients and a linear $f(Q)=\alpha Q$ model with $\alpha \neq 0$. We present the best fit values of the viscosity coefficients and the model parameter by using the combined H(z)+Pantheon+BAO data set. Then we analyze the asymptotic behavior of our model using the dynamical system approach. Further, we investigate the behavior of effective pressure, effective EoS parameter, and the statefinder parameters. In Sec \ref{sec5}, we consider a non-linear $f(Q)$ model, specifically, $f(Q)=Q+\beta Q^2$ and then investigated its dynamical behavior. In the last section \ref{sec6}, we present the outcomes of our investigation.

\section{Geometry with Non-metricity}\label{sec2}
\justify

It is well known that gravitational interactions in the standard theory of gravity (GR) ruled by the curvature tensor,
\begin{equation}\label{2a}
R^\alpha_{\: \beta\mu\nu} = 2\partial_{[\mu} X^\alpha_{\: \nu]\beta} + 2X^\alpha_{\: [\mu \mid \lambda \mid}X^\lambda_{\nu]\beta}
\end{equation} 
Here the affine connection $X^\alpha_{\: \mu\nu}= \Gamma^\alpha_{\: \beta\gamma}$ represents torsion-free metric compatible Levi-Civita connection that is characterized by the metric tensor. We start by recalling that the tool by which gravity is mediated is not the physical manifold, but the generic affine connection \cite{Tom}. The generic affine connection can be described through other properties, such as non-metricity. According to the strong equivalence principle \cite{CMS}, corresponding to every point on the physical manifold, there is well defined tangent space and this affine connection act as a mediator between neighboring tangent spaces to define the derivative operators. This implies that the generic affine connection follows a decomposition \cite{Tom},
\begin{equation}\label{2b}
	X^\alpha_{\ \mu\nu}=\Gamma^\alpha_{\ \mu\nu}+K^\alpha_{\ \mu\nu}+L^\alpha_{\ \mu\nu},
\end{equation}
where  
\begin{equation}\label{2c}
	\Gamma^\alpha_{\ \mu\nu}\equiv\frac{1}{2}g^{\alpha\lambda}(g_{\mu\lambda,\nu}+g_{\lambda\nu,\mu}-g_{\mu\nu,\lambda})
\end{equation}
denotes the Levi-Civita connection of the metric tensor $g_{\mu\nu}$,
\begin{equation}\label{2d}
K^\alpha_{\ \mu\nu}\equiv\frac{1}{2}(T^{\alpha}_{\ \mu\nu}+T_{\mu \ \nu}^{\ \alpha}+T_{\nu \ \mu}^{\ \alpha})
\end{equation}
is the contortion tensor, and
\begin{equation}\label{2e}
L^\alpha_{\ \mu\nu}\equiv\frac{1}{2}(Q^{\alpha}_{\ \mu\nu}-Q_{\mu \ \nu}^{\ \alpha}-Q_{\nu \ \mu}^{\ \alpha})
\end{equation}	
represents the distortion tensor.
Here we define the torsion tensor $T^\alpha_{\ \mu\nu}$ and the non-metricity tensor $Q_{\alpha\mu\nu}$ as 
\begin{equation}\label{2f}
 T^\alpha_{\ \mu\nu}\equiv X^\alpha_{\ \mu\nu}-X^\alpha_{\ \nu\mu}
\end{equation}
and 
\begin{equation}\label{2g}
Q_{\alpha\mu\nu}\equiv\nabla_\alpha g_{\mu\nu}
\end{equation} 
We impose the symmetric teleparallelism condition $R^\alpha_{\: \beta\mu\nu} = 0$ and $K^\alpha_{\ \mu\nu} = 0$ to develop $f(Q)$ theory. The condition of symmetric teleparallelism makes the generic affine connection to be total inertial. In an arbitrary coordinate system, this connection can be parameterized as \cite{jimenez/2020}
\begin{equation}\label{2h}
X^\alpha \,_{\mu \beta} = \frac{\partial x^\alpha}{\partial \xi^\rho} \partial_ \mu \partial_\beta \xi^\rho.
\end{equation}
We can always choose a coordinate transformation so that $ X^\alpha_{\ \mu\nu} = 0 $. This special coordinate frame is known as coincident gauge.

The action corresponding to $f(Q)$ gravity under the geometrical framework that incorporates the connection to be flat and symmetric reads as \cite{JLL}
\begin{equation}\label{2i}
S= \int{\frac{1}{2}f(Q)\sqrt{-g}d^4x} + \int{L_m\sqrt{-g}d^4x}
\end{equation}
Here $f(Q)$ represents the function of the non-metricity scalar $Q$, $L_m$ is the matter Lagrangian, and $g=det( g_{\mu\nu} )$.
From the symmetry of the metric tensor, it follows that one can construct two different non-metricity vectors
\begin{equation}\label{2j}
Q_\alpha = Q_\alpha\:^\mu\:_\mu \: and\:  \tilde{Q}_\alpha = Q^\mu\:_{\alpha\mu}
\end{equation}
In addition, the superpotential tensor is defined by
\begin{equation}\label{2k}
4P^\lambda\:_{\mu\nu} = -Q^\lambda\:_{\mu\nu} + 2Q_{(\mu}\:^\lambda\:_{\nu)} + (Q^\lambda - \tilde{Q}^\lambda) g_{\mu\nu} - \delta^\lambda_{(\mu}Q_{\nu)}.
\end{equation}
One can acquire the non-metricity scalar by using superpotential tensor as \cite{LZ} 
\begin{equation}\label{2l}
Q = -Q_{\lambda\mu\nu}P^{\lambda\mu\nu}. 
\end{equation}
Further, the energy momentum tensor characterizes the matter-energy content of the universe reads as
\begin{equation}\label{2m}
\mathcal{T}_{\mu\nu} = \frac{-2}{\sqrt{-g}} \frac{\delta(\sqrt{-g}L_m)}{\delta g^{\mu\nu}}
\end{equation}
By altering the action \eqref{2i} with respect to the metric, we obtained the field equation
\begin{widetext}
\begin{equation}\label{2n}
\frac{2}{\sqrt{-g}}\nabla_\lambda (\sqrt{-g}f_Q P^\lambda\:_{\mu\nu}) + \frac{1}{2}g_{\mu\nu}f+f_Q(P_{\mu\lambda\beta}Q_\nu\:^{\lambda\beta} - 2Q_{\lambda\beta\mu}P^{\lambda\beta}\:_\nu) = -T_{\mu\nu}.
\end{equation}
\end{widetext}

In the absence of hypermomentum, we obtained the following field equation by altering the action \eqref{2i} with respect to the connection,
\begin{equation}\label{2o}
\nabla_\mu \nabla_\nu (\sqrt{-g}f_Q P^{\mu\nu}\:_\lambda) =  0 
\end{equation}
Moreover, it follows from the Bianchi identity that this equation is fulfilled automatically once the metric equations of motion are \cite{LAR}.

\section{Flat FLRW Universe in Symmetric Teleparallel Cosmology with Bulk Viscous Matter}\label{sec3}
\justify
We begin with following spatially flat FLRW line element \cite{Ryden} in
Cartesian coordinates, which is, as a matter of fact also a coincident gauge coordinates, therefore from now connection becomes trivial and metric is only a fundamental variable,
 \begin{equation}\label{3a}
ds^2= -dt^2 + a^2(t)[dx^2+dy^2+dz^2]
\end{equation}
where, $ a(t) $ is the scale factor that measures the cosmic expansion. However, the most general connection in cosmological settings was derived in \cite{e9,FDA}, which leads to a non-trivial contribution to the field equations. Now, for the line element \eqref{3a} we obtain the non-metricity scalar $Q$ as
\begin{equation}\label{3b}
 Q= 6H^2  
\end{equation}
The energy momentum tensor that corresponds to the universe with bulk viscous matter for the line element \eqref{3a} reads as,  
\begin{equation}\label{3c}
\mathcal{T}_{\mu\nu}=(\rho+\bar{p})u_\mu u_\nu + \bar{p}g_{\mu\nu}
\end{equation}
Here $\rho$ represents matter-energy density, $\bar{p}=p-3\zeta H$ is the effective pressure having coefficient of bulk viscosity $\zeta$ and the usual pressure $p$, and $u^\mu=(1,0,0,0)$ are components of the four velocities.\\
The field equations characterizing the universe with bulk viscous matter are\cite{LZ}, 
\begin{equation}
3H^{2}=\frac{1}{2f_{Q}}\left( -\rho +\frac{f}{2}\right)  \label{3d}
\end{equation}
and 
\begin{equation}
\dot{H}+3H^{2}+\frac{\dot{f_{Q}}}{f_{Q}}H=\frac{1}{2f_{Q}}\left( \bar{p}+
\frac{f}{2}\right) \label{3e}
\end{equation}
In addition, we acquire the following matter conservation equation by taking the trace of the field equations,
\begin{equation}\label{3f}
\dot{\rho} + 3H\left(\rho+\bar{p}\right)=0
\end{equation}

In further analysis we consider only the dust case, so the effective pressure is given purely by viscous pressure $\bar{p}=-3\zeta H$.

\section{Linear $f(Q)$ model }\label{sec4}
\justify
We consider the following $f(Q)$ function for our analysis \cite{RS,ZHH},  
\begin{equation}\label{3g}
f(Q)=\alpha Q,\ \ \ \alpha \neq 0  
\end{equation}
Then the Friedmann equations for this specific $f(Q)$ function becomes, 
\begin{equation}\label{3h}
\rho =-3\alpha H^{2}  
\end{equation}
and 
\begin{equation}\label{3i}
\bar{p}=2\alpha \dot{H}+3\alpha H^{2} 
\end{equation}
In particular, for the case $\alpha=-1$, one can retrieve the usual Friedmann equations of GR.

In fluid mechanics, it is well known that the coefficient of bulk viscosity is associated with the rate of expansion or compression of the fluid \cite{DZ1}. In the context of the cosmological model, the cosmic fluid is comoving with the expanding universe, therefore the velocity $\dot{a}$ and acceleration $\ddot{a}$ of the expanding universe and that of the cosmic fluid becomes coincide. Thus it is evident that the bulk viscosity coefficient $\zeta$ is proportional to the velocity and acceleration term. We consider the following bulk viscosity form, which is nothing but a linear combination of velocity and acceleration terms with a constant \cite{DZ2}, 
\begin{equation}\label{3j}
\zeta =\zeta_{0}+\zeta_{1}\left( \frac{\dot{a}}{a}\right) +\zeta_{2}\left( \frac{{
\ddot{a}}}{\dot{a}}\right) =\zeta_{0}+\zeta_{1}H+\zeta_{2}\left( \frac{\dot{H}}{
H}+H\right)  
\end{equation}
Now we set $\bar{\zeta_{0}}$, $\bar{\zeta_{1}}$, and $\bar{\zeta_{2}}$ as
\begin{equation}\label{3k}
\frac{3\zeta _{0}}{H_{0}}=\bar{\zeta_{0}}, \ \ 3\zeta_{1}=\bar{\zeta_{1}}, \ \ 3\zeta_{2}=
\bar{\zeta_{2}}
\end{equation}
where $H_{0}$ is the Hubble parameter value at present time $z=0$ and this parameters are known as dimensionless bulk viscous parameters.
Then by using \eqref{3j} and \eqref{3k} in equation \eqref{3i} and the fact that $\frac{d}{dt}=H\frac{d}{dln(a)}$, we have 
\begin{equation} \label{3l}
\frac{dH}{dln(a)}+\left( \frac{3\alpha +\bar{\zeta_{1}}+ \bar{\zeta_{2}}}{2\alpha +\bar{\zeta_{2}}}\right) H+\left( \frac{\bar{\zeta_{0}}}{2\alpha +\bar{\zeta_{2}}}\right)H_{0}=0
\end{equation}
Now by integrating the above equation, we obtained the following expression of Hubble parameter in terms of redshift corresponding to the bulk viscous non-relativistic matter dominated universe 
\begin{widetext}
\begin{equation}\label{3m}
H(z)=H_{0}\left[ (1+z)^{\left( \frac{3\alpha +\bar{\zeta_{1}}+ \bar{\zeta_{2}}}{2\alpha +\bar{
\zeta_{2}}}\right) }\left( 1+\frac{\bar{\zeta_{0}}}{3\alpha +\bar{\zeta_{1}}+ \bar{\zeta_{2}}}
\right) -\frac{\bar{\zeta_{0}}}{3\alpha +\bar{\zeta_{1}}+ \bar{\zeta_{2}}}\right] 
\end{equation}
\end{widetext}
Now we consider the following three different cases on dimensionless viscous parameters which is well known in the literature, 

\justify Case I : When viscosity coefficient depends on both the velocity and acceleration i.e. $\bar{\zeta_{0}}$, $\bar{\zeta_{1}}$, and $\bar{\zeta_{2}}$ all are non-zero.\\
Case II: When viscosity coefficient depends on the velocity but not on acceleration i.e. $\bar{\zeta_{0}}$, $\bar{\zeta_{1}}$ are non-zero whereas $\bar{\zeta_{2}}=0$. In this case, the expression for the Hubble parameter becomes
\begin{equation}\label{3n}
H(z)=H_{0}\left[ (1+z)^{\left( \frac{3\alpha +\bar{\zeta_{1}}}{2\alpha} \right) }\left( 1+\frac{\bar{\zeta_{0}}}{3\alpha +\bar{\zeta_{1}}} \right) -\frac{\bar{\zeta_{0}}}{3\alpha +\bar{\zeta_{1}}} \right] 
\end{equation}
Case III: When viscosity coefficient does not depends on both the velocity and acceleration i.e.  $\bar{\zeta_{1}}=0$, $\bar{\zeta_{2}}=0$ whereas $\bar{\zeta_{0}}$ is non-zero. In this case, the expression for the Hubble parameter becomes
\begin{equation}\label{3o}
H(z)=H_{0}\left[ (1+z)^{\frac{3}{2}} \left( 1+\frac{\bar{\zeta_{0}}}{3\alpha} \right) -\frac{\bar{\zeta_{0}}}{3\alpha } \right] 
\end{equation}
In particular, when $\bar{\zeta_{0}}=\bar{\zeta_{1}}=\bar{\zeta_{2}}=0$, then $H(z)$ reduces to $H(z)=H_{0}(1+z)^{\frac{3}{2}}$ that corresponds to the non-viscous matter dominated universe.

\subsection{Parameter Estimation Using Observational Data}\label{sec4a}
\justify
In this section, we consider updated H(z) data, Pantheon Supernovae data, and the Baryonic Acoustic Oscillation (BAO) data to estimate the parameter values corresponding to all three different cases. We apply the Markov Chain Monte Carlo (MCMC) approach along with Bayesian analysis to explore the parameter space of our cosmological bulk viscous model by utilizing \texttt{emcee} python library \cite{Mackey/2013}.

\subsubsection{H(z) datasets}
\justify
We are familiar with the fact that the Hubble parameter can directly predict the rate of cosmic expansion. In general, there are two very popular approaches to extracting the Hubble parameter value at definite redshift, namely differential age and line of sight BAO technique. In this manuscript, we work with an updated list of H(z) data points. One can check the reference for the complete set of data points \cite{RS}. Furthermore, we have taken $H_0=67.9$ Km/s/Mpc \cite{Planck} for our investigation. 
The $\chi^2$ function corresponding to H(z) data points reads as
\begin{equation}\label{4a}
\chi_{H}^{2}=\sum\limits_{k=1}^{57}
\frac{[H_{th}(z_{k},\theta)-H_{obs}(z_{k})]^{2}}{
\sigma _{H(z_{k})}^{2}}.  
\end{equation}
Here, $H_{obs}$ is the Hubble parameter value extracted from the cosmic observations while $H_{th}$ represents its theoretical value calculated at $z_{k}$ with parameter space $\theta$, and  $\sigma_{H(z_{k})}$ denotes corresponding error.

\subsubsection{Pantheon datasets}
\justify
In the present manuscript, we work with a recently published Pantheon supernovae data set that contains 1048 supernovae samples with their distance moduli $\mu^{obs}$ in the redshift range $z \in [0.01, 2.3]$ \cite{Scolnic/2018}. The $\chi^2$ function corresponding to Pantheon data points reads as
\begin{equation}\label{4b}
\chi^2_{SN}=\sum_{i,j=1}^{1048}\bigtriangledown\mu_{i}\left(C^{-1}_{SN}\right)_{ij}\bigtriangledown\mu_{j},
\end{equation}
Here $C_{SN}$ is the covariance matrix \cite{Scolnic/2018}, and
\begin{align*}\label{4c}
\quad \bigtriangledown\mu_{i}=\mu^{th}(z_i,\theta)-\mu_i^{obs}.
\end{align*} 
is the difference between the observed value of distance modulus extracted from the cosmic observations and its theoretical values calculated from the model with given parameter space $\theta$. We define the distance modulus by $\mu=m_B-M_B$, where $m_B$ and $M_B$ denote respectively the observed apparent magnitude and the absolute magnitude at a given redshift (Retrieving the nuisance parameter following the recent approach called BEAMS with Bias Correction (BBC) \cite{BMS}). Furthermore, its theoretical value is given by
\begin{equation}\label{4d}
\mu(z)= 5log_{10} \left[ \frac{D_{L}(z)}{1 Mpc}  \right]+25, 
\end{equation}
where 
\begin{equation}\label{4e}
D_{L}(z)= c(1+z) \int_{0}^{z} \frac{ dx}{H(x,\theta)}
\end{equation}

\subsubsection{BAO datasets}
\justify
Baryonic Acoustic Oscillation (BAO) investigates oscillations produced in the early phase of the universe due to cosmological perturbations in the fluid consisting of photons, baryons, and dark matter, which is tightly coupled through Thompson scattering. The BAO measurements consist of Sloan Digital Sky Survey (SDSS), Six Degree Field Galaxy Survey (6dFGS), and the Baryon Oscillation spectroscopic Survey (BOSS) \cite{BAO1,BAO2}. The relations used in BAO measurements are,
\begin{equation}\label{4f}
d_{A}(z)=\int_{0}^{z}\frac{dz^{\prime }}{H(z^{\prime })},
\end{equation}
\begin{equation}\label{4g}
D_{V}(z)=\left( d_{A}(z)^{2}z/H(z) \right)^{1/3},
\end{equation}
and
\begin{equation}\label{4h}
\chi _{BAO}^{2}=X^{T}C^{-1}X  
\end{equation}
where $C$ is the covariance matrix \cite{BAO6}, $d_{A}(z)$ denotes the angular diameter distance while $D_{V}(z)$ represents the dilation scale.

\subsubsection{Observational Results}
\justify
We obtained the constraints on parameters of our cosmological model with bulk viscous matter corresponding to all three different cases for the combined H(z)+Pantheon+BAO data set by minimizing the total chi-squared function $\chi_{H}^{2}+\chi_{SN}^{2}+\chi_{BAO}^{2}$.

\begin{widetext}
\begin{center}
\begin{figure}[H]
\includegraphics[scale=0.95]{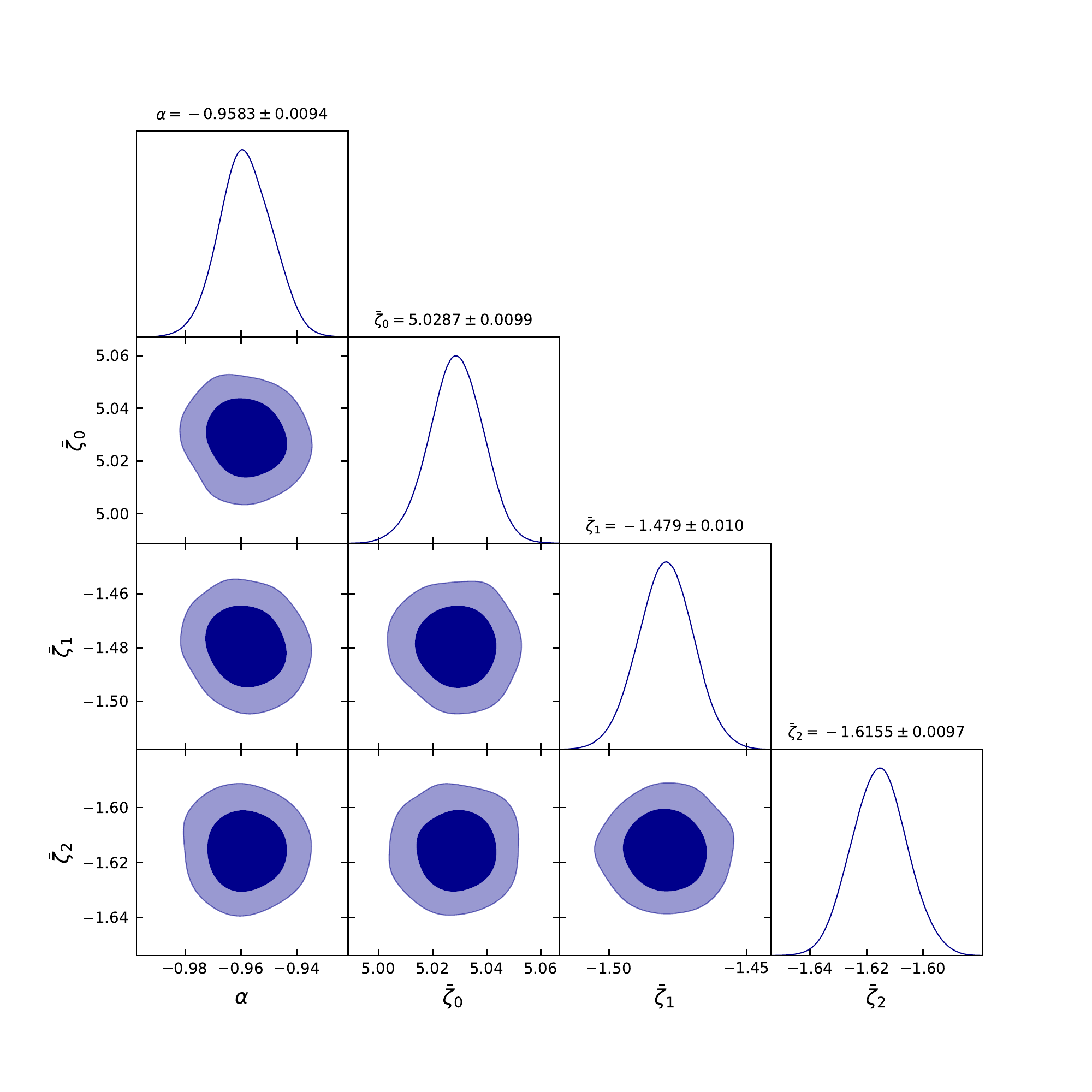}
\caption{Constraints on the model and bulk viscous parameters corresponding to Case I at $1-\sigma$ and $2-\sigma$ confidence interval using the combined H(z)+Pantheon+BAO data set.}\label{f1}
\end{figure}
\end{center}
\end{widetext}

\begin{widetext}
\begin{center}
\begin{figure}[h]
\includegraphics[scale=0.65]{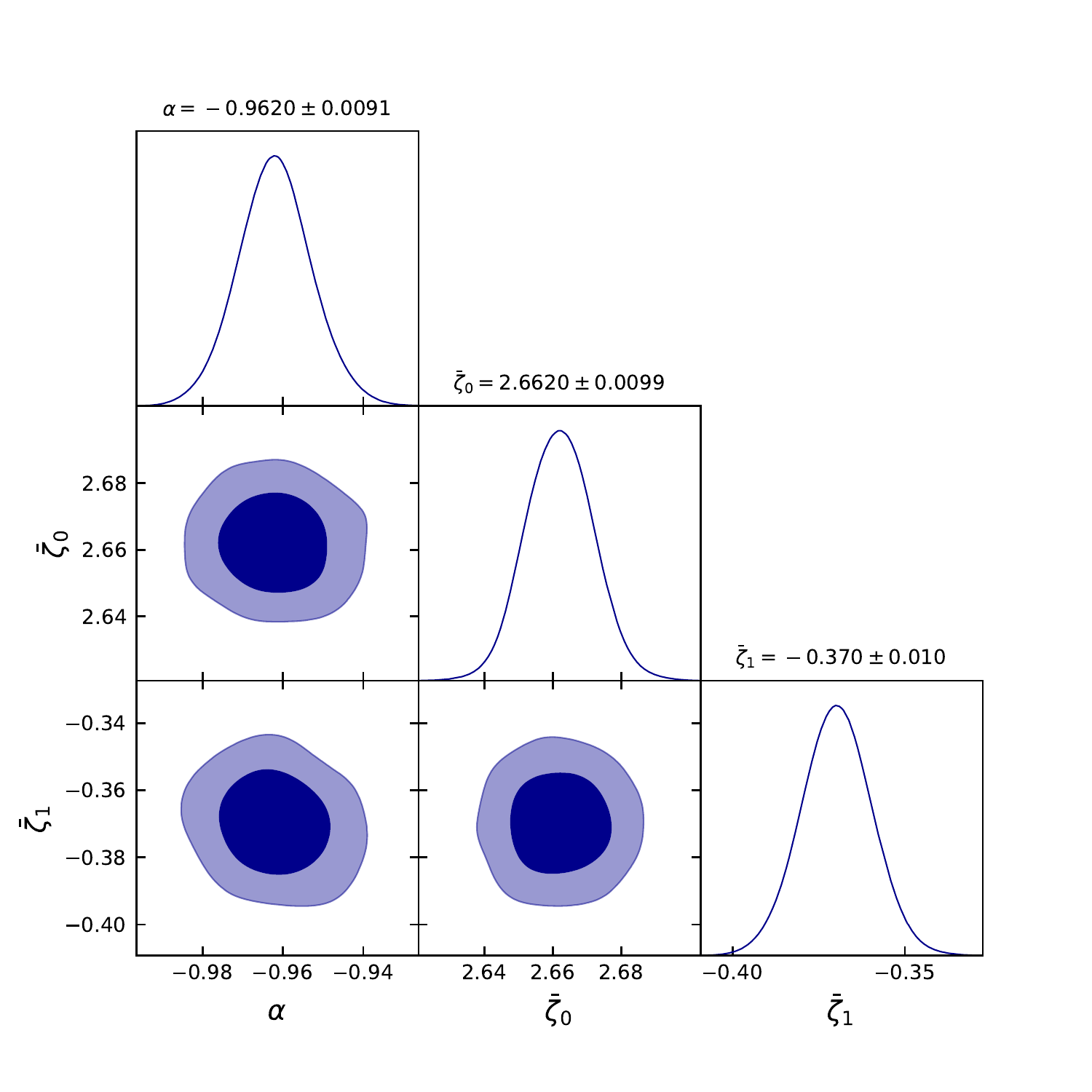}
\caption{Constraints on the model and bulk viscous parameters corresponding to Case II at $1-\sigma$ and $2-\sigma$ confidence interval using the combined H(z)+Pantheon+BAO data set.}\label{f2}
\end{figure}
\end{center}
\end{widetext}

\begin{widetext}
\begin{center}
\begin{figure}[h]
\includegraphics[scale=0.85]{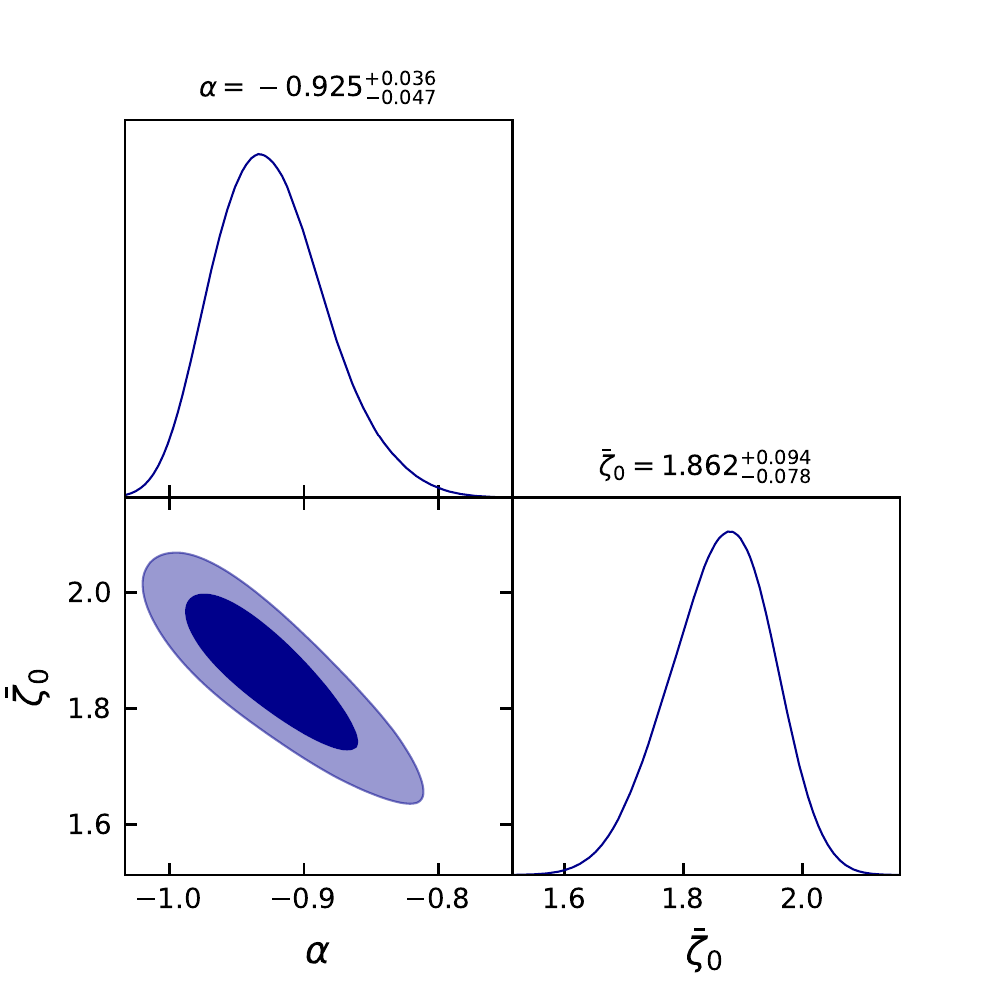}
\caption{Constraints on the model and bulk viscous parameters corresponding to Case III at $1-\sigma$ and $2-\sigma$ confidence interval using the combined H(z)+Pantheon+BAO data set.}\label{f3}
\end{figure}
\end{center}
\end{widetext}

The obtained constraints on the model and bulk viscous parameters corresponds to all three cases for the combined H(z)+Pantheon+BAO data set presented in Table \ref{Table-1}.

\begin{widetext}
\begin{table}[H]
\begin{center}\caption{Table shows the constraints on the model and bulk viscous parameters corresponds to all three cases for the combined H(z)+Pantheon+BAO data set.}
\begin{tabular}{|c|c|c|c|c|}
\hline
Cases & $\alpha$ & $\bar{\zeta}_0$  & $\bar{\zeta}_1$ & $\bar{\zeta}_2$ \\
\hline 
Case I & $-0.9583 \pm 0.0094$ & $5.0287 \pm 0.0099$ & $-1.479 \pm 0.010$ & $-1.6155 \pm 0.0097$ \\
\hline
Case II & $-0.9620 \pm 0.0091$ & $2.6620 \pm 0.0099$ & $-0.370 \pm 0.010$ & $0$ \\
\hline
Case III & $-0.925^{+0.036}_{-0.047} $ & $1.862^{+0.094}_{-0.078} $ & $0$ & $0$ \\
\hline
\end{tabular}\label{Table-1}
\end{center}
\end{table}
\end{widetext}

\subsection{Phase Space Analysis}\label{sec4b}
\justify
In this section, we are going to investigate the asymptotic behavior of our cosmological bulk viscous model by utilizing the dynamical system approach. First, we convert the cosmological equations of our model into a set of autonomous differential equations, and then we analyze the corresponding phase space. Now, we define the following dimensionless variables,
\begin{equation}\label{5a}
x= \frac{-\rho}{3\alpha H^2}  \:\: \text{and} \:\:   y=\frac{1}{\frac{H_0}{H}+1}
\end{equation}
The variables $x$ and $y$ are called phase space variables and it lies in the range $0 \leqslant y \leqslant 1$, whereas the Friedmann equation \eqref{3h} implies $ x = 1$.
\justify \textbf{Case I : When viscosity coefficient depends on both the velocity and acceleration i.e. $\zeta =\zeta_{0}+\zeta_{1}\left( \frac{\dot{a}}{a}\right) +\zeta_{2}\left( \frac{{\ddot{a}}}{\dot{a}}\right) =\zeta_{0}+\zeta_{1}H+\zeta_{2}\left( \frac{\dot{H}}{H}+H\right)$}.

\justify We start with defining the variable $N=ln(a)$. Then by solving Friedmann equations with conservation equation for the bulk viscous matter, we obtained the following autonomous differential equations in the variables $x$ and $y$, 
\begin{widetext}
\begin{equation}\label{5b}
x'=\frac{dx}{dN}=\frac{(x-1)}{(2\alpha+\bar{\zeta}_2)y} \left[ 2\bar{\zeta}_0 (1-y)+(2\bar{\zeta}_1-\bar{\zeta}_2)y \right] = F_1(x,y)
\end{equation}
\begin{equation}\label{5c}
y'=\frac{dy}{dN}=\frac{(y-1)}{(2\alpha+\bar{\zeta}_2)} \left[ \bar{\zeta}_0 (1-y)+(\bar{\zeta}_1+\bar{\zeta}_2+3\alpha)y \right] = F_2(x,y)
\end{equation}
\end{widetext}
Further, by using the definition of the equation of state and deceleration parameter, we acquired
\begin{equation}\label{5d}
q=\frac{1}{(2\alpha+\bar{\zeta}_2)} \left[ \alpha+ \bar{\zeta}_1+ \frac{\bar{\zeta}_0 (1-y)}{y} \right]    
\end{equation}
and
\begin{equation}\label{5e}
\omega= \frac{1}{3(2\alpha+\bar{\zeta}_2)} \left[ 2\bar{\zeta}_1 - \bar{\zeta}_2 + \frac{2\bar{\zeta}_0 (1-y)}{y} \right]
\end{equation}
Now by solving equations $x'=0$ and $y'=0$, we obtained the coordinates of the critical points $(x_c,y_c)$ corresponding to autonomous equations \eqref{5b} and \eqref{5c} as,
\begin{equation}\label{5f}
(x_c,y_c)=(1,1) \:\: \text{and} \:\: (x_c,y_c)=(1,\frac{\bar{\zeta}_0}{\bar{\zeta}_0-\bar{\zeta}_1-\bar{\zeta}_2-3\alpha})
\end{equation}
We investigate the stability of the given autonomous system in the neighbourhood of the critical points. First we linearize the given autonomous dynamical system by assuming small perturbations near the critical points $(x,y)\longrightarrow (x_c+\delta x,y_c+\delta y)$ satisfying

\begin{equation*}\label{5g}
\left[
\begin{array}{c}
\delta x'  \\ 
\delta y'  \\ 
\end{array}
\right] \,=   \left[
\begin{array}{cc}
\left( \frac{\partial F_1}{\partial x} \right)_{0} & \left( \frac{\partial F_1}{\partial y}  \right)_{0}  \\ 
\left( \frac{\partial F_2}{\partial x}  \right)_{0} & \left( \frac{\partial F_2}{\partial x}  \right)_{0} \\ 
\end{array}
\right]  \left[
\begin{array}{c}
\delta x \\ 
\delta y  \\ 
\end{array}
\right] 
\end{equation*}
Here the suffix 0 implies that the above Jacobian matrix calculated at the critical points $(x_c,y_c)$ and it is given as
\begin{widetext}
\begin{equation}\label{5h}
J= \left[
\begin{array}{cc}
\left( \frac{1}{(2\alpha+\bar{\zeta}_2)y} \left[ 2\bar{\zeta}_0 (1-y)+(2\bar{\zeta}_1-\bar{\zeta}_2)y \right] \right)_{0} & \left( \frac{-2\bar{\zeta}_0 (x-1)}{(2\alpha+\bar{\zeta}_2)y^2}  \right)_{0}  \\ 
0 & \left( \frac{1}{(2\alpha+\bar{\zeta}_2)} \left[ -2\bar{\zeta}_0 (y-1)+(\bar{\zeta}_1+\bar{\zeta}_2+3\alpha)(2y-1) \right] \right)_{0} \\ 
\end{array}
\right] 
\end{equation}
\end{widetext}
Now we analyze the behavior of each of the critical points obtained in \eqref{5f}.
\justify \textbf{(i)} $(x_c,y_c)=(1,1)$ : \\
In this case the eigenvalues obtained for the Jacobian matrix $J$ are 
\begin{equation}\label{5i}
\lambda_1= \frac{(2\bar{\zeta}_1-\bar{\zeta}_2)}{(2\alpha+\bar{\zeta}_2)}  \:\: \text{and} \:\: \lambda_2= \frac{(\bar{\zeta}_1+\bar{\zeta}_2+3\alpha)}{(2\alpha+\bar{\zeta}_2)}
\end{equation}
By using the constrained values of the parameters, one can have
\begin{equation}\label{5j}
\lambda_1= 0.38 \:\: \text{and} \:\: \lambda_2= 1.69
\end{equation}
Since $\lambda_1>0$ and $\lambda_2>0$, therefore the critical point $(1,1)$ is unstable. Further, $y_c=1$ implies $\frac{1}{\frac{H_0}{H}+1}=1$ i.e. $\frac{H_0}{H}=0$ which indicates that either $H_0=0$ or $H \rightarrow \infty $. Since $H_0 \neq 0$, therefore the critical point $(x_c,y_c)=(1,1)$ represents initial singularity characterized by $H \rightarrow \infty $ i.e. a past attractor. Moreover from equations \eqref{5d} and \eqref{5e}, we obtained $q \sim 0.69$ and $\omega \sim 0.12$.

\justify \textbf{(ii)} $(x_c,y_c)=(1,\frac{\bar{\zeta}_0}{\bar{\zeta}_0-\bar{\zeta}_1-\bar{\zeta}_2-3\alpha})=(1,0.4572)$ : \\
In this case the eigenvalues obtained for the Jacobian matrix $J$ are 
\begin{equation}\label{5k}
\lambda_1= -3  \:\: \text{and} \:\: \lambda_2= -\frac{(\bar{\zeta}_1+\bar{\zeta}_2+3\alpha)}{(2\alpha+\bar{\zeta}_2)}
\end{equation}
Again by using the constrained values of the parameters, we can have
\begin{equation}\label{5l}
\lambda_1= -3 \:\: \text{and} \:\: \lambda_2= -1.69
\end{equation}
Since $\lambda_1<0$ and $\lambda_2<0$, therefore the critical point $(1,0.4572)$ is stable. In addition, from equations \eqref{5d} and \eqref{5e}, we found that $q \sim -1$ and $\omega \sim -1$. Hence the critical point $(x_c,y_c)=(1,0.4572)$ corresponds to de-Sitter phase representing a future attractor.

\begin{widetext}
\begin{table}[H]
\begin{center}\caption{Table shows the critical points and their behavior corresponding to Case I.}
\begin{tabular}{|c|c|c|c|c|}
\hline
Critical Points $(x_c,y_c)$ & Eigenvalues $\lambda_1$ and $\lambda_2$ & Nature of critical point  & $q$ & $\omega$ \\
\hline 
$(1,1)$ & $0.38 \:\: \text{and} \:\: 1.69$ & Unstable & $0.69$ & $0.12$ \\
\hline
$(1,0.4572)$ & $-3 \:\: \text{and} \:\: -1.69$ & Stable & $-1$ & $-1$ \\
\hline
\end{tabular}\label{Table-2}
\end{center}
\end{table}
\end{widetext}

\begin{figure}[h]
\includegraphics[scale=0.6]{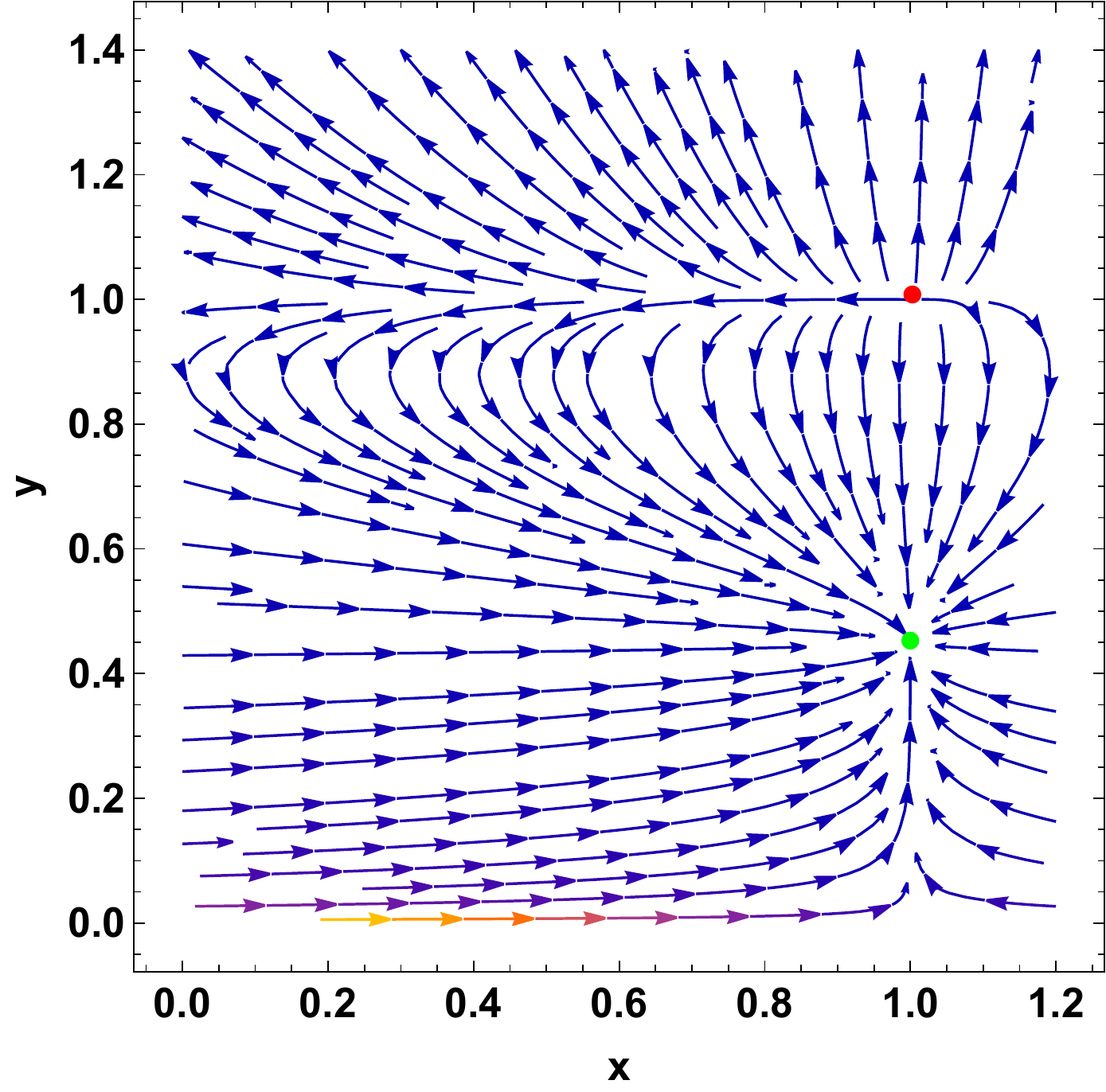}
\caption{Phase plot in the $x-y$ plane corresponding to Case I with a red dot and green dot denoting the past and future attractor, respectively, followed by the arrowhead representing the direction of the trajectories.  }\label{f4}
\end{figure}
\justify From the phase plot presented in the Fig \eqref{f4} it is evident that the evolutionary trajectory of our bulk viscous cosmological model emerges from the critical point $(1,1)$ as a past attractor and then it converges to the critical point $(1,0.4572)$ that is nothing but a future attractor. Thus our model corresponding to Case I depicts the evolution of the universe starting with an initial singularity and behaves like a de-Sitter phase in the far future.

\justify \textbf{Case II : When viscosity coefficient depends on the velocity but not on acceleration i.e.  $\zeta =\zeta_{0}+\zeta_{1}\left( \frac{\dot{a}}{a}\right)  =\zeta_{0}+\zeta_{1}H$}.

\justify In this case, the autonomous system of equations \eqref{5b} and \eqref{5c} becomes
\begin{equation}\label{5m}
x'=\frac{(x-1)}{\alpha y} \left[ \bar{\zeta}_0 (1-y)+\bar{\zeta}_1 y \right] 
\end{equation}
\begin{equation}\label{5n}
y'=\frac{(y-1)}{2\alpha} \left[ \bar{\zeta}_0 (1-y)+(\bar{\zeta}_1+3\alpha)y \right]
\end{equation}
The deceleration and equation of state parameter given in \eqref{5d} and \eqref{5e} reduces to
\begin{equation}\label{5o}
q=\frac{1}{2\alpha} \left[ \alpha+ \bar{\zeta}_1+ \frac{\bar{\zeta}_0 (1-y)}{y} \right]    
\end{equation}
and
\begin{equation}\label{5p}
\omega= \frac{1}{6\alpha} \left[ 2\bar{\zeta}_1 + \frac{2\bar{\zeta}_0 (1-y)}{y} \right]
\end{equation}
Now on solving equations $x'=0$ and $y'=0$, we obtained the coordinates of the critical points $(x_c,y_c)$ corresponding to autonomous equations \eqref{5m} and \eqref{5n} as,
\begin{equation}\label{5q}
(x_c,y_c)=(1,1) \:\: \text{and} \:\: (x_c,y_c)=(1,\frac{\bar{\zeta}_0}{\bar{\zeta}_0-\bar{\zeta}_1-3\alpha})
\end{equation}
We investigate the behavior of each of the critical points obtained in \eqref{5q}.
\justify \textbf{(i)} $(x_c,y_c)=(1,1)$ : \\
In this case the eigenvalues obtained for the reduced Jacobian matrix $J$ are 
\begin{equation}\label{5r}
\lambda_1= \frac{\bar{\zeta}_1}{\alpha} \sim 0.38  \:\: \text{and} \:\: \lambda_2= \frac{(\bar{\zeta}_1+3\alpha)}{2\alpha} \sim 1.69
\end{equation}
Since $\lambda_1>0$ and $\lambda_2>0$, therefore the critical point $(1,1)$ is unstable. In addition, as in the previous case, $y_c=1$ implies either $H_0=0$ or $H \rightarrow \infty $ and hence the critical point $(x_c,y_c)=(1,1)$ represents initial singularity characterized by $H \rightarrow \infty $ i.e. a past attractor since $H_0$ is non-zero. Further from equations \eqref{5o} and \eqref{5p}, we obtained $q \sim 0.69$ and $\omega \sim 0.12$.

\justify \textbf{(ii)} $(x_c,y_c)=(1,\frac{\bar{\zeta}_0}{\bar{\zeta}_0-\bar{\zeta}_1-3\alpha})=(1,0.4498)$ : \\
In this case the eigenvalues obtained for the reduced Jacobian matrix $J$ are 
\begin{equation}\label{5s}
\lambda_1= -3  \:\: \text{and} \:\: \lambda_2= -\frac{(\bar{\zeta}_1+3\alpha)}{2\alpha} \sim -1.69
\end{equation}
Since $\lambda_1<0$ and $\lambda_2<0$, therefore the critical point $(1,0.4498)$ is stable. From equations \eqref{5o} and \eqref{5p}, we obtained $q \sim -1$ and $\omega \sim -1$. Hence the critical point $(x_c,y_c)=(1,0.4498)$ corresponds to de-Sitter type universe representing a future attractor.

\begin{widetext}
\begin{table}[H]
\begin{center}\caption{Table shows the critical points and their behavior corresponding to Case II.}
\begin{tabular}{|c|c|c|c|c|}
\hline
Critical Points $(x_c,y_c)$ & Eigenvalues $\lambda_1$ and $\lambda_2$ & Nature of critical point  & $q$ & $\omega$ \\
\hline 
$(1,1)$ & $0.38 \:\: \text{and} \:\: 1.69$ & Unstable & $0.69$ & $0.12$ \\
\hline
$(1,0.4498)$ & $-3 \:\: \text{and} \:\: -1.69$ & Stable & $-1$ & $-1$ \\
\hline
\end{tabular}\label{Table-3}
\end{center}
\end{table}
\end{widetext}

\begin{figure}[H]
\includegraphics[scale=0.6]{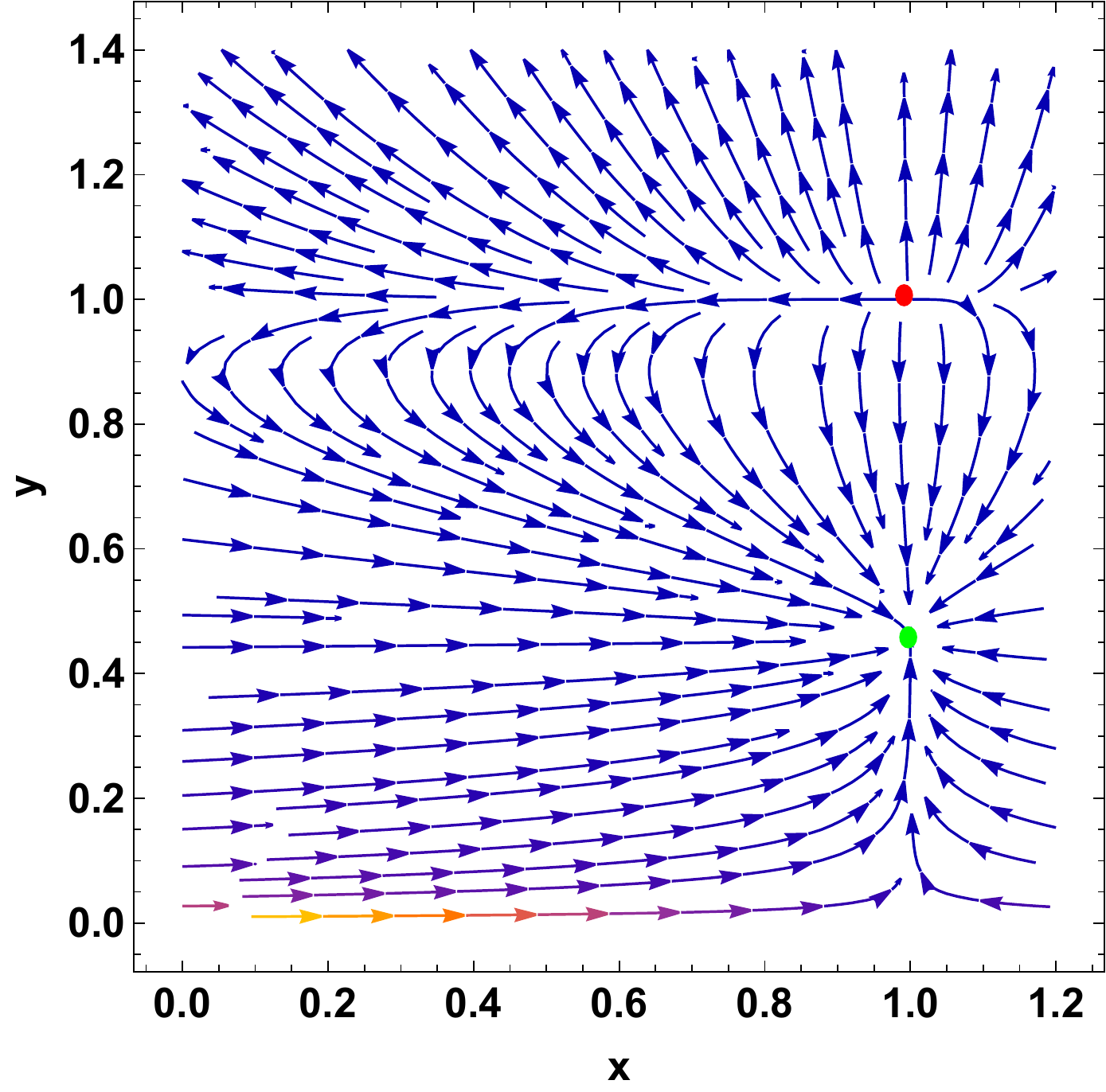}
\caption{Phase plot in the $x-y$ plane corresponding to Case II with a red dot and green dot denoting the past and future attractor, respectively, followed by the arrowhead representing the direction of the trajectories.  }\label{f5}
\end{figure}
\justify From the phase plot presented in the Fig \eqref{f5} it is clear that the evolutionary trajectory of our model emerges from the critical point $(1,1)$ and then it converges to the critical point $(1,0.4498)$. Thus our model with bulk viscosity corresponding to Case II represents the universe evolving from an initial singularity to de-Sitter type universe in the far future, as we obtained in Case I.

\justify \textbf{Case III :  When viscosity coefficient does not depends on both the velocity and acceleration i.e.  $\zeta =\zeta_{0}$}.

\justify In this case, the autonomous system of equations \eqref{5b} and \eqref{5c} becomes
\begin{equation}\label{5t}
x'=\frac{\bar{\zeta}_0(x-1)(1-y)}{\alpha y }  
\end{equation}
\begin{equation}\label{5u}
y'=\frac{(y-1)}{2\alpha} \left[ \bar{\zeta}_0 (1-y)+3\alpha y \right]
\end{equation}
The deceleration and equation of state parameter given in \eqref{5d} and \eqref{5e} reduces to
\begin{equation}\label{5v}
q=\frac{1}{2} + \frac{\bar{\zeta}_0 (1-y)}{2\alpha y} 
\end{equation}
and
\begin{equation}\label{5w}
\omega=  \frac{\bar{\zeta}_0(1-y)}{3\alpha y} 
\end{equation}
On solving equations $x'=0$ and $y'=0$, we obtained the coordinates of the critical points $(x_c,y_c)$ corresponding to autonomous equations \eqref{5t} and \eqref{5u} as,
\begin{equation}\label{5x}
(x_c,y_c)=(x,1) \:\: \text{and} \:\: (x_c,y_c)=(1,\frac{\bar{\zeta}_0}{\bar{\zeta}_0-3\alpha})
\end{equation}
\justify \textbf{(i)} $(x_c,y_c)=(x,1)$ : \\
In this case the eigenvalues obtained for the reduced Jacobian matrix $J$ are 
\begin{equation}\label{5y}
\lambda_1= 0  \:\: \text{and} \:\: \lambda_2= \frac{3}{2}
\end{equation}
It is evident that first component $x$ of the critical point varies from 0 to 1 whereas $y_c=1$.  Therefore the critical point $(x_c,y_c)$ is not an isolated point rather it is a line of critical points and that indicates initial state of universe since $H \rightarrow \infty $. As $\lambda_1=0$ and $\lambda_2>0$, therefore every critical point $(x,1)$ on the line are unstable i.e. a past attractor. Further we obtained $q=\frac{1}{2} $ and $\omega =0$ by using equations \eqref{5v} and \eqref{5w}. These values of the equation of state and deceleration parameter corresponds to a decelerated matter dominated universe.  

\justify \textbf{(ii)} $(x_c,y_c)=(1,\frac{\bar{\zeta}_0}{\bar{\zeta}_0-3\alpha})=(1,0.4015)$ : \\
In this case the eigenvalues obtained for the reduced Jacobian matrix $J$ are 
\begin{equation}\label{5z}
\lambda_1= -3  \:\: \text{and} \:\: \lambda_2= -\frac{3}{2} 
\end{equation}
Since $\lambda_1<0$ and $\lambda_2<0$, therefore the critical point $(1,0.4015)$ is stable. From equations \eqref{5v} and \eqref{5w}, we obtained $q \sim -1$ and $\omega \sim -1$. Hence the critical point $(x_c,y_c)=(1,0.4015)$ corresponds to de-Sitter type universe representing a future attractor.

\begin{widetext}
\begin{table}[H]
\begin{center}\caption{Table shows the critical points and their behavior corresponding to Case III.}
\begin{tabular}{|c|c|c|c|c|}
\hline
Critical Points $(x_c,y_c)$ & Eigenvalues $\lambda_1$ and $\lambda_2$ & Nature of critical point  & $q$ & $\omega$ \\
\hline 
$(x,1)$ & $0 \:\: \text{and} \:\: \frac{3}{2}$ & Unstable & $\frac{1}{2}$ & $0$ \\
\hline
$(1,0.4015)$ & $-3 \:\: \text{and} \:\: -\frac{3}{2}$ & Stable & $-1$ & $-1$ \\
\hline
\end{tabular}\label{Table-4}
\end{center}
\end{table}
\end{widetext}

\begin{figure}[H]
\includegraphics[scale=0.6]{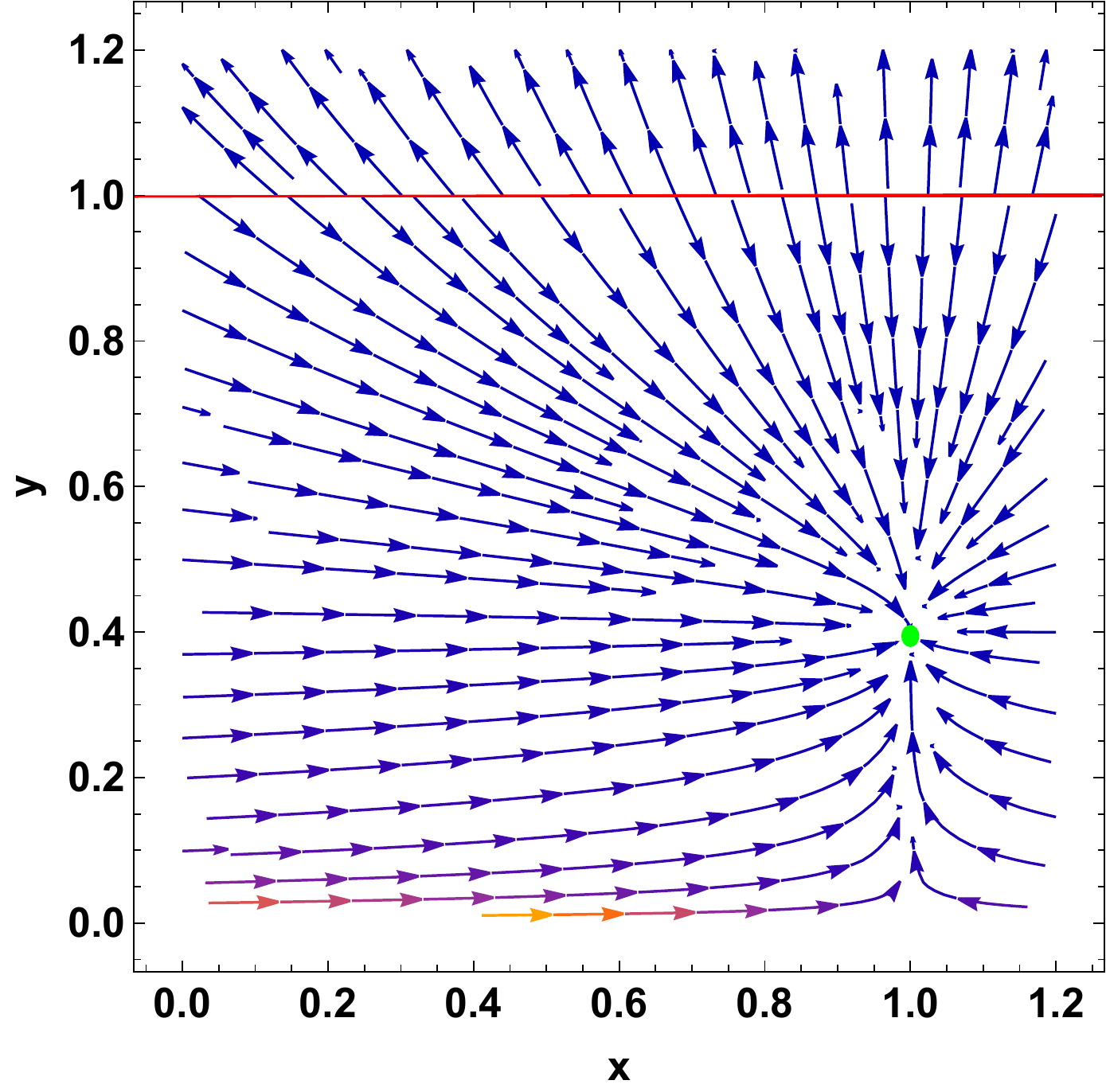}
\caption{Phase plot in the $x-y$ plane corresponding to Case III with the red line of critical points and green dot denoting the past and future attractor, respectively, followed by the arrowhead representing the direction of the trajectories.  }\label{f6}
\end{figure}
\justify From the phase plot presented in the Fig \eqref{f6} it is clear that the evolution trajectory of our model emerges from the non isolated critical point $(x,1)$ that corresponds to a decelerated matter dominated universe and then it converges to the critical point $(1,0.4015)$ representing a future attractor. Thus our cosmological model with bulk viscosity behaves like $\Lambda$CDM model. We found that the accelerated deSitter like phase comes purely from the $\bar{\zeta}_0$ case without any geometrical modification to GR. Further, due to the constraint $x=1$, only the critical point $(1,1)$ lie on the line of critical points $(x,1)$ is physical.

\subsection{Behavior of Cosmological Parameters}\label{sec4c}
\justify
In this section we present physical behavior of well known cosmological parameters such as pressure component in the presence of viscosity, effective equation of state (EoS), and the $r-s$ parameter \cite{V.S.}.

\begin{figure}[H]
\includegraphics[scale=0.5]{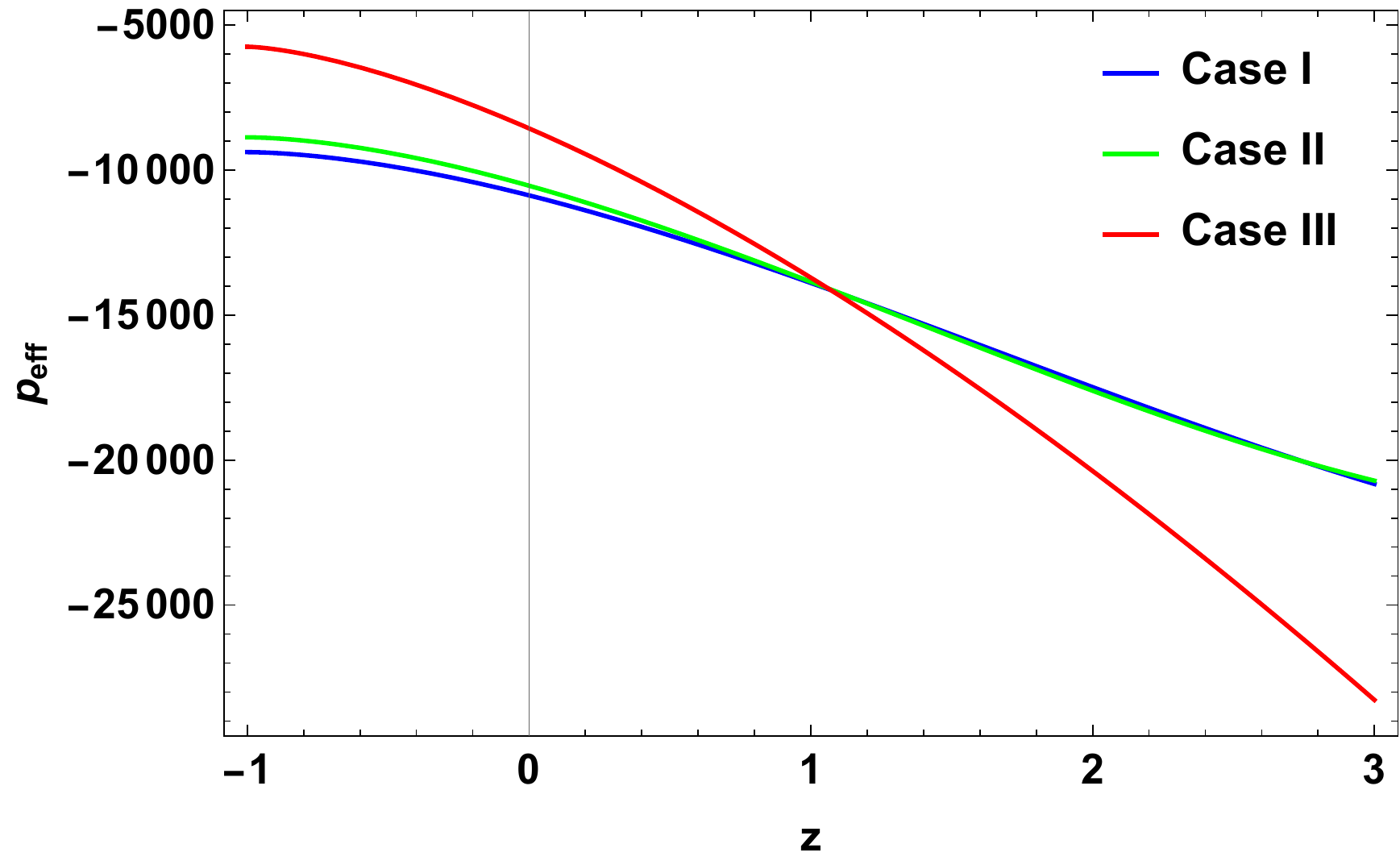}
\caption{Plot for pressure component in the presence of bulk viscosity corresponding to the observational constraints obtained using combined H(z)+Pantheon+BAO data set.}\label{f7}
\end{figure}

\begin{figure}[H]
\includegraphics[scale=0.5]{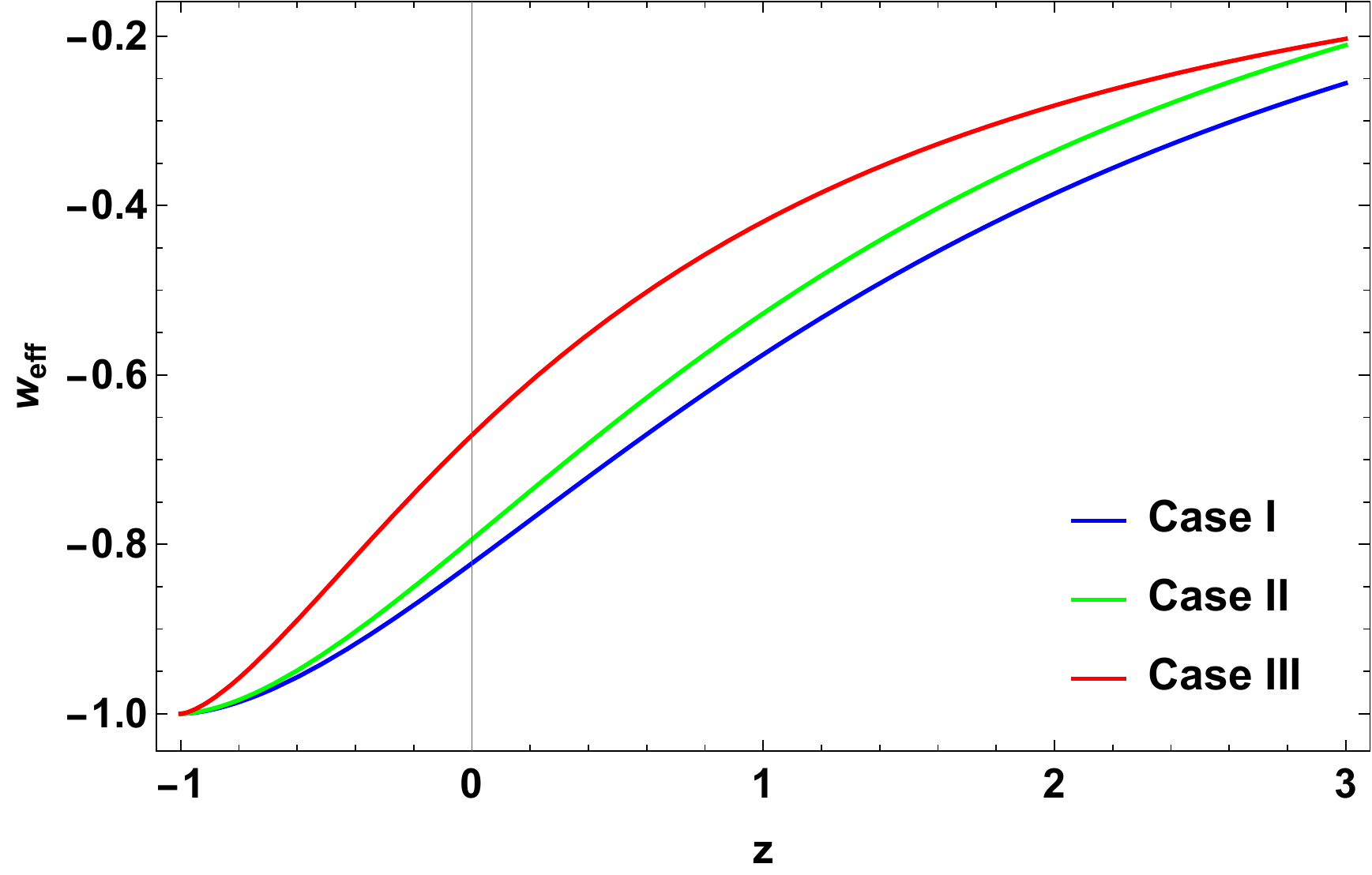}
\caption{Plot for effective EoS parameter corresponding to the observational constraints obtained using combined H(z)+Pantheon+BAO data set.}\label{f8}
\end{figure}

\begin{figure}[H]
\includegraphics[scale=0.62]{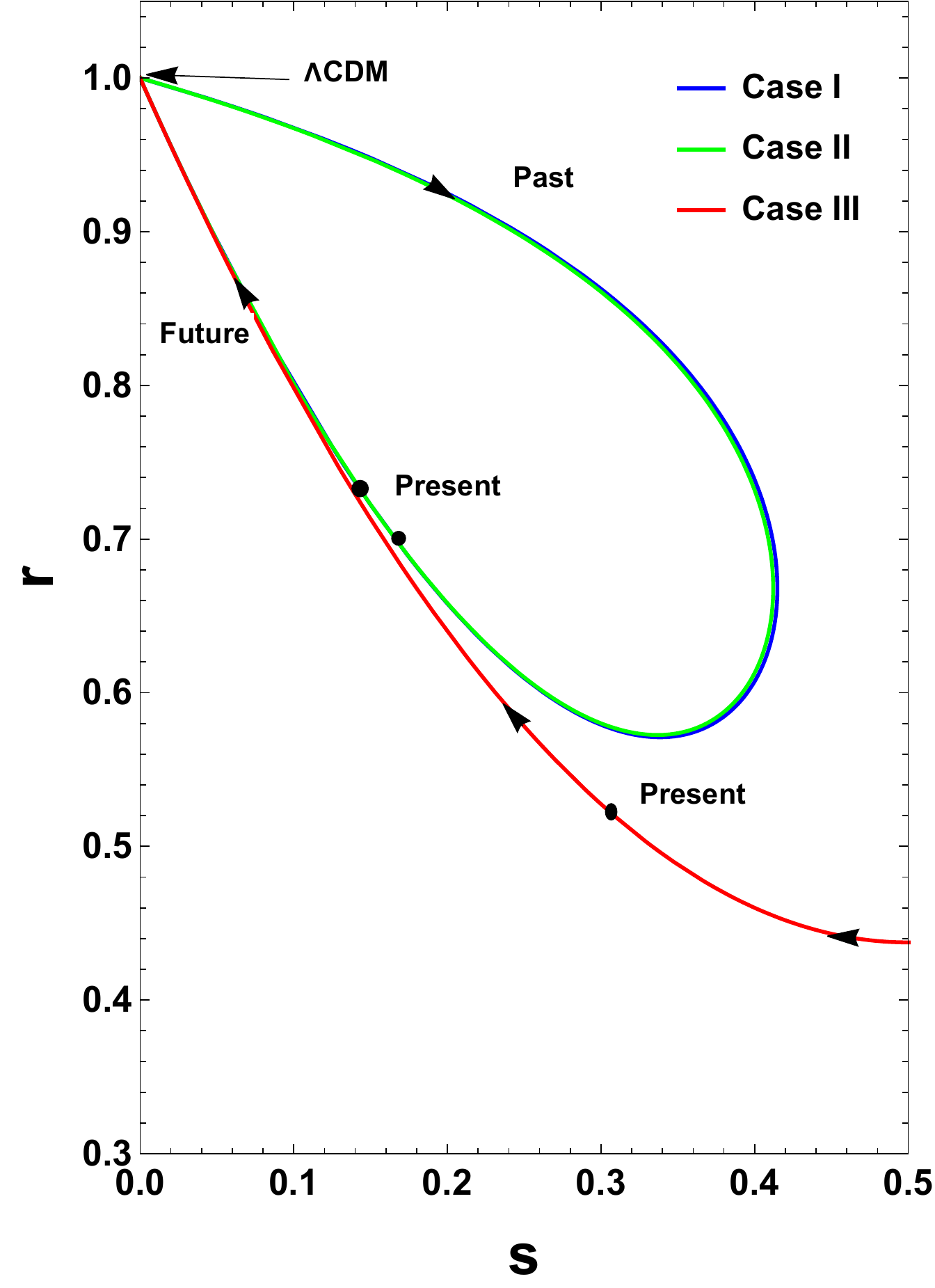}
\caption{Plot for $r-s$ parameter corresponding to the observational constraints obtained using combined H(z)+Pantheon+BAO data set.}\label{f9}
\end{figure}

\justify Fig\eqref{f7} indicates that the pressure component in the presence of bulk viscosity shows negative behavior for all three cases. The negative behavior of the pressure component confirms the existence of a mysterious dark energy component leading to acceleration. The plot for the effective EoS parameter in fig\eqref{f8} supports the recently observed acceleration, and it will converge to the $\Lambda$CDM equation of state in the far future. In fig\eqref{f9}, we present the evolutionary trajectory of the given viscous fluid model in the $r-s$ plane corresponding to all three cases. Since all three trajectories lie in the region $r<1$ and $s>0$, our bulk viscous model follows the quintessence scenario. Further, we observed that corresponding to all three cases, the trajectories of our model converge to the $\Lambda$CDM fixed point. Thus, from the $r-s$ diagram, we conclude that our viscous fluid cosmological model represents a stable de-Sitter phase of the universe in the far future.

\section{Non-Linear $f(Q)$ model }\label{sec5}
\justify

We consider the following $f(Q)$ function for our analysis \cite{jimenez/2020},
\begin{equation}\label{51}
f(Q)= -Q+\beta Q^2 
\end{equation}
We consider only the case $\bar{p}=-3\zeta_{0}H= -\bar{\zeta}_0 H_0 H$. Then by using Friedmann equations \eqref{3d} and \eqref{3e} we obtained the following first-order non-linear differential equation,
\begin{equation}\label{52}
2\dot{H} (36\beta H^2-1)+3H^2 (18\beta H^2-1) + \bar{\zeta}_0 H_0 H =0 
\end{equation}
Now, we define the following dimensionless variables,
\begin{equation}\label{53}
x= \frac{-\rho}{3H^2(18 \beta H^2-1)}  \:\: \text{and} \:\:   y=\frac{1}{\frac{H_0}{H}+1}
\end{equation}
The variable $y$ lies in the range $0 \leqslant y \leqslant 1$ and from equation \eqref{3d}, we have constraint $x=1$.
Then we obtained the following autonomous differential equations corresponding to our non-linear $f(Q)$ model, 
\begin{widetext}
\begin{equation}\label{54}
x'=\frac{dx}{dN}=  \frac{\bar{\zeta}_0 (x-1)(1-y)^3}{y \left[ \bar{\beta}y^2-(1-y)^2 \right] } 
\end{equation}
\begin{equation}\label{55}
y'=\frac{dy}{dN}=\frac{(y-1)}{2\left[2 \bar{\beta}y^2-(1-y)^2 \right]} \left[ 3y \lbrace \bar{\beta}y^2-(1-y)^2 \rbrace + \bar{\zeta}_0 (1-y)^3 \right]
\end{equation}
\end{widetext}
Here $\bar{\beta}=18H_0^2 \beta $. Further, by using the definition of the equation of state and deceleration parameter, we acquired
\begin{widetext}
\begin{equation}\label{57}
q=-1+ \frac{1}{2y\left[2 \bar{\beta}y^2-(1-y)^2 \right]} \left[ 3y \lbrace \bar{\beta}y^2-(1-y)^2 \rbrace + \bar{\zeta}_0 (1-y)^3 \right]
\end{equation}
\textbf{and}
\begin{equation}\label{58}
\omega= -1+ \frac{1}{3y\left[2 \bar{\beta}y^2-(1-y)^2 \right]} \left[ 3y \lbrace \bar{\beta}y^2-(1-y)^2 \rbrace + \bar{\zeta}_0 (1-y)^3 \right]
\end{equation}
\end{widetext}

Since $\bar{\zeta}_0>0$, we fix  $\bar{\zeta}_0 =1$ and then we have investigated the stability of the autonomous system given by \eqref{54}-\eqref{55} in the neighbourhood of the critical points. In particular, for $\bar{\beta}=0$, this autonomous system is same as that of Case III and hence the dynamics. Therefore we analyzed our non-linear $f(Q)$ model for the parameter values $\bar{\beta}=-1$ and $\bar{\beta}=1$. The obtained results are presented in the Table \ref{Table-5}.

\begin{widetext}
\begin{table}[H]
\begin{center}\caption{Table shows the critical points and their behavior corresponding to non-linear $f(Q)$ model.}
\begin{tabular}{|c|c|c|c|c|c|}
\hline
& Critical Points $(x_c,y_c)$ & Eigenvalues $\lambda_1$ and $\lambda_2$ & Nature of critical point  & $q$ & $\omega$ \\
\hline 
$\bar{\beta}=-1$&$\left( x,1 \right)$ & $\frac{3}{4}\:\: \text{and} \:\: 0$ & Unstable & $-\frac{1}{4}$ & $-\frac{1}{2}$ \\
&$\left( 1,0.234 \right)$ & $-3\:\:  \text{and} \:\: -1.62$ & Stable & $-1$ & $-1$ \\
\hline
& $\left(x,1\right)$ & $\frac{3}{4}\:\: \text{and} \:\: 0$ & Unstable & $-\frac{1}{4}$ & $-\frac{1}{2}$ \\
$\bar{\beta}=1$ & $\left( 1,0.283 \right)$ & $-3\:\:  \text{and} \:\: -1.15$ & Stable & $-1$ & $-1$ \\
& $\left( 1,0.426 \right)$ & $-3\:\:  \text{and} \:\: -9.6$ & Stable & $-1$ & $-1$ \\
\hline
\end{tabular}\label{Table-5}
\end{center}
\end{table}
\end{widetext}

\begin{figure}[H]
\includegraphics[scale=0.75]{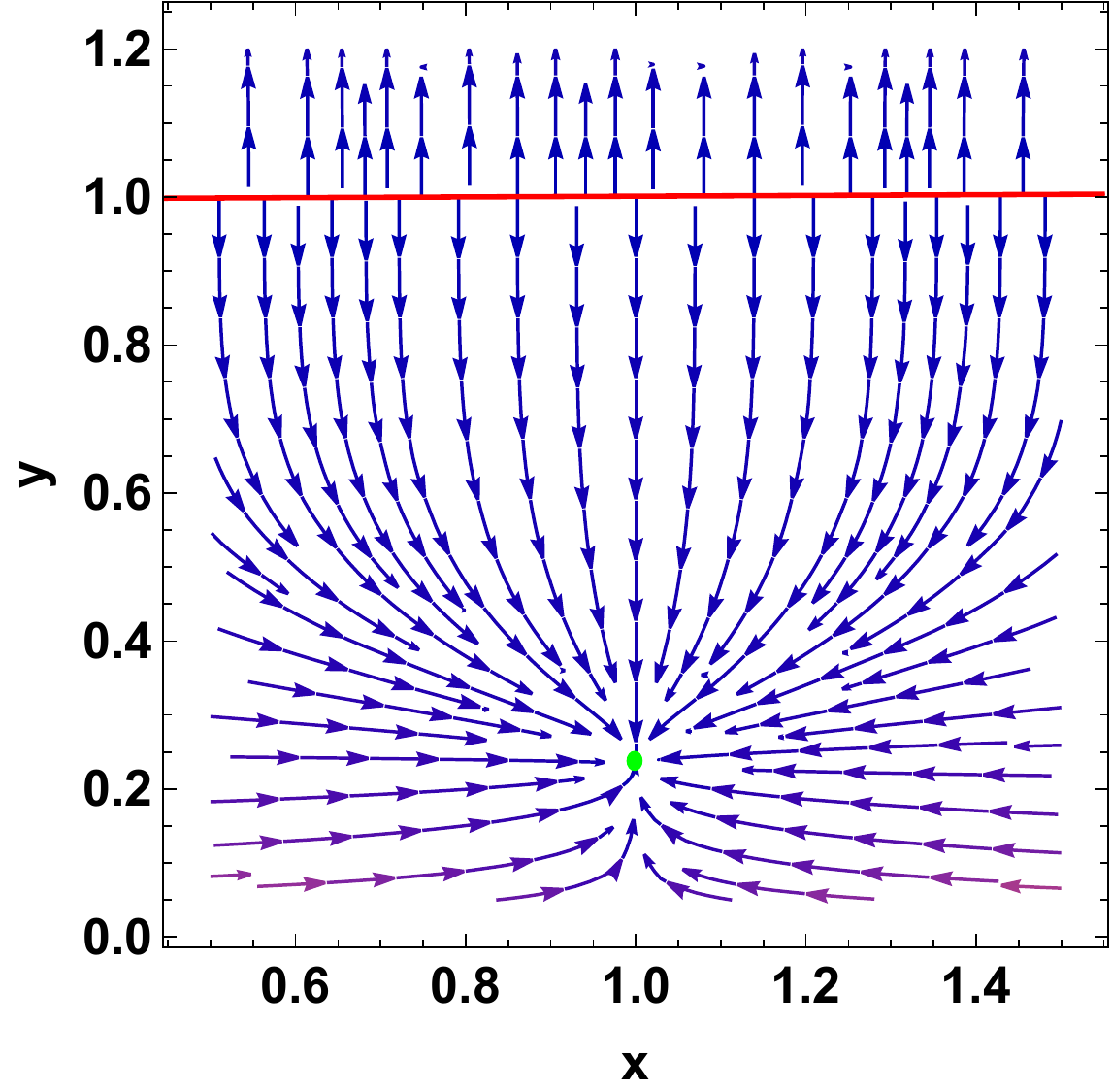}
\caption{Phase plot in the $x-y$ plane corresponding to the non-linear $f(Q)$ model ($\bar{\beta}=-1$) with the red line of critical points and green dot denoting the past and future attractor, respectively, followed by the arrowhead representing the direction of the trajectories.  }\label{n1}
\end{figure}

\begin{figure}[H]
\includegraphics[scale=0.71]{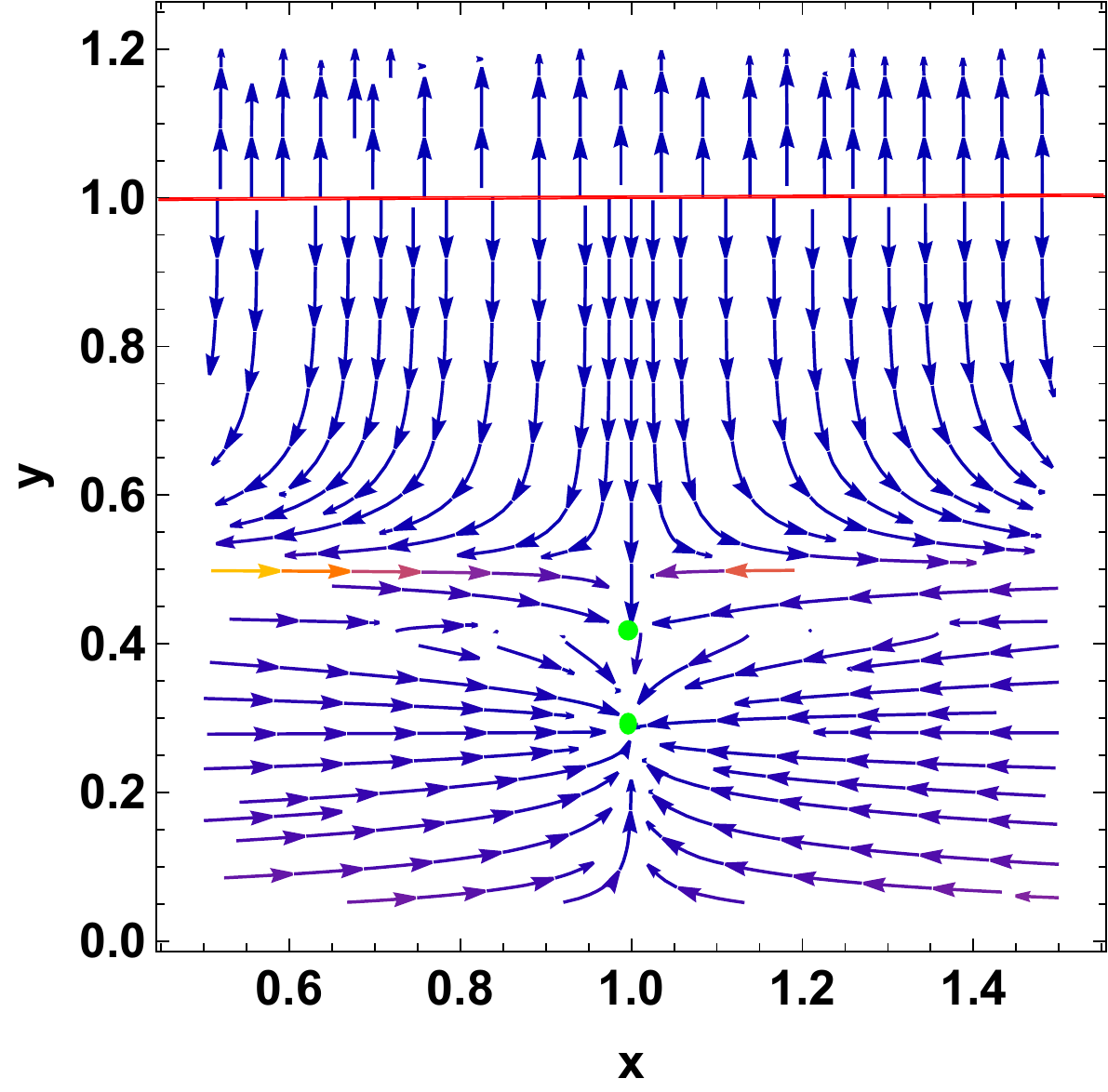}
\caption{Phase plot in the $x-y$ plane corresponding to the non-linear $f(Q)$ model ($\bar{\beta}=1$) with the red line of critical points and green dot denoting the past and future attractor, respectively, followed by the arrowhead representing the direction of the trajectories.  }\label{n2}
\end{figure}

\begin{figure}[H]
\includegraphics[scale=0.48]{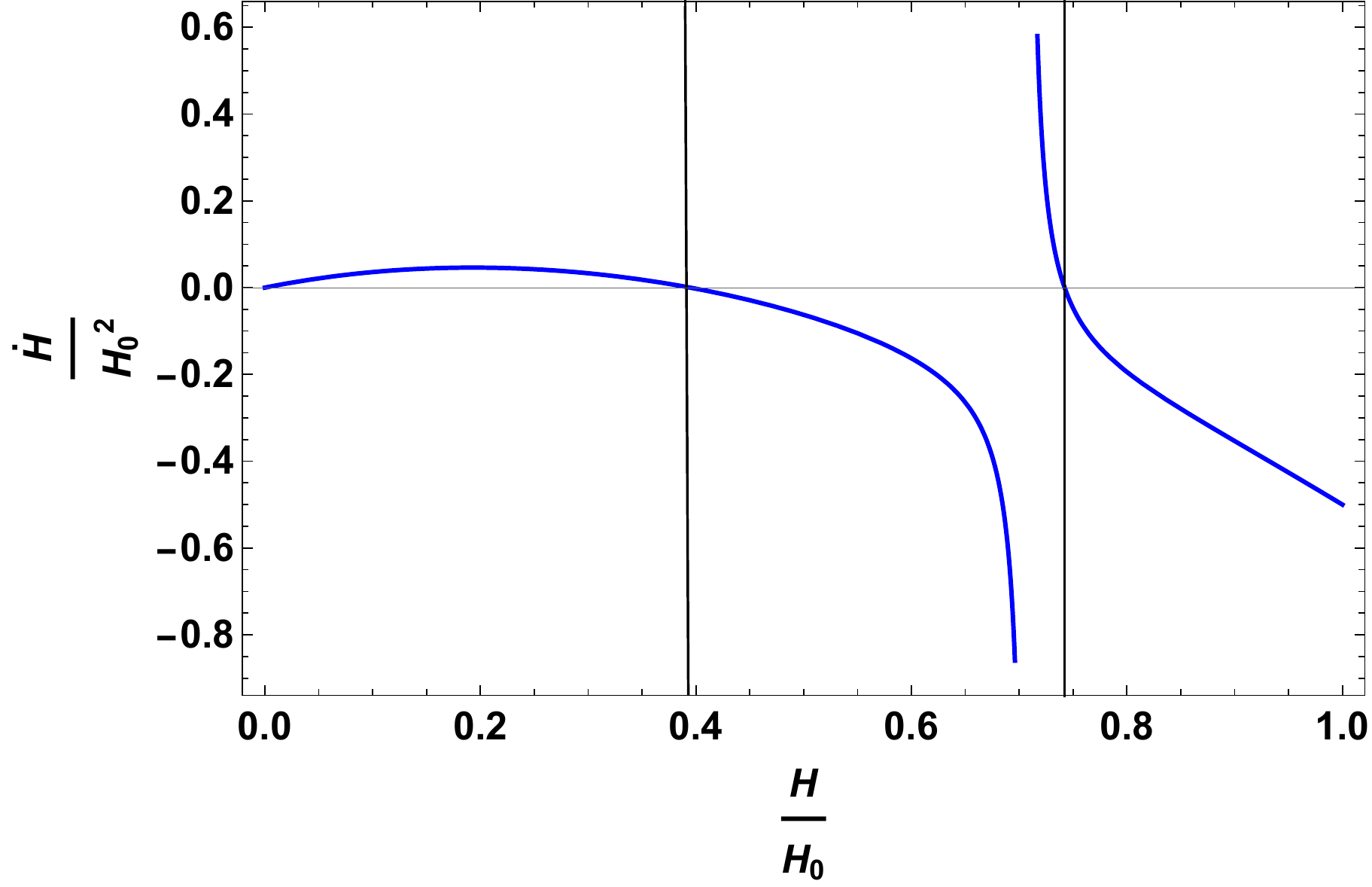}
\caption{$\dot{H}$ as a function of $\frac{H}{H_0}$ from equation \eqref{52} with $\bar{\beta}=\bar{\zeta}_0=1$, where vertical black lines indicating the fixpoints that corresponds to $\dot{H}=0$.  }\label{add}
\end{figure}

From the Table \ref{Table-5} it is clear that there is only one de-Sitter type future attractor for the case $\bar{\beta}=-1$ whereas two future attractors corresponding to $\bar{\beta}=1$ case. From the phase plot presented in the Fig \eqref{n1} it is clear that the evolution trajectory of our non-linear model, for the case $\bar{\beta}=-1$, emerges from the non isolated critical points $(x,1)$ and then it converges to the critical point $(1,0.234)$ representing a de-Sitter like accelerated phase. Further note that, due to the constraint $x=1$, only the critical point $(1,1)$ lie on the line of critical points $(x,1)$ is physical. Moreover, from the phase plot presented in the Fig \eqref{n2} we found that the evolution trajectory of our non-linear model, for the case $\bar{\beta}=1$, emerges from the critical point $(1,1)$ and then settles down to two stable critical points $(1,0.283)$ and $(1,0.426)$ representing the de-Sitter like accelerated phase. The behavior of $\dot{H}$ as a function of $\frac{H}{H_0}$ from equation \eqref{52} with $\bar{\beta}=\bar{\zeta}_0=1$ is presented in Fig \eqref{add}. We can observe that two regimes are separated by a line  $H=\frac{H_0}{\sqrt{2\bar{\beta}}}$ of diverging $\dot{H}$ and the evolution only to the upper attractor $\left( 1,0.426 \right)$ from an initial big bang $y=1$ is possible. Thus, only the fixed point $\left( 1,0.426 \right)$ is a viable attractor from an initial big bang. Hence, evolution from the big bang $H \rightarrow \infty$ can only occur towards the right fix point at $H \approx 0.75 H_0$.

\section{Conclusion}\label{sec6}
\justify
In this manuscript, we have investigated the role of different bulk viscosity coefficients in cosmic evolution under the framework of symmetric teleparallel gravity. We have considered three different cases of bulk viscosity coefficients which is well known in the literature, namely $(i)\zeta =\zeta_{0}+\zeta_{1}\left( \frac{\dot{a}}{a}\right) +\zeta_{2}\left( \frac{{\ddot{a}}}{\dot{a}}\right) $, $(ii)\zeta =\zeta_{0}+\zeta_{1}\left( \frac{\dot{a}}{a}\right)$, and $(iii)\zeta =\zeta_{0}$. Then we calculated Hubble parameter value in terms of redshift corresponding to all three cases by assuming a linear $f(Q)$ model, specifically, $f(Q)=\alpha Q$ where $\alpha \neq 0$.  We presented the observation constraints (see Table \ref{Table-1} and Fig\eqref{f1}-\eqref{f3}) on the model parameter and viscosity coefficients corresponding to all the three cases using the combined H(z)+Pantheon+BAO data set. Then we investigated the asymptotic behavior of our cosmological bulk viscous model by using phase space analysis. We derived the set of autonomous differential equations and the corresponding critical points for all three cases (see Table \ref{Table-2},\ref{Table-3}, and \ref{Table-4}). Further we presented the phase space plot corresponding to all the three cases (see Fig\eqref{f4},\eqref{f5}, and \eqref{f6}). We found that our viscous fluid model represents a universe evolving from matter dominated decelerated epoch, which is a past attractor, to a stable de-sitter accelerated epoch, which is a future attractor. In addition, we investigated the physical behavior of the pressure component in the presence of viscosity, the effective equation of state (EoS), and the $r-s$ parameter. We found that the pressure component in the presence of bulk viscosity presented in fig\eqref{f7} shows negative behavior for all three cases. The plot for effective EoS parameter presented in fig\eqref{f8} shows that the current universe is going through a period of accelerated expansion. Lastly, from fig\eqref{f9}, we obtained that trajectories of our cosmological viscous model lie in the quintessence region. Further, these trajectories converge to the $\Lambda$CDM fixed point for all three cases, which coincides with results obtained in the phase space analysis. We found that the accelerated deSitter like phase comes purely from the $\bar{\zeta}_0$ case without any geometric modification to GR. Moreover, we found that the late-time behavior of all three cases of viscosity coefficients are identical. Further, we have considered a non-linear $f(Q)$ model, specifically, $f(Q)=-Q+\beta Q^2$ and then we analyzed the behavior of model using dynamical approach presented in Table \ref{Table-5}. We found that there is only one de-Sitter type future attractor for the case $\bar{\beta}=-1$ whereas two future attractors corresponding to $\bar{\beta}=1$ case (see Fig\eqref{n1} and \eqref{n2}). However, the upper attractor $\left( 1,0.426 \right)$ in Fig \eqref{n2} is a viable attractor from an initial big bang, and therefore evolution from the big bang $H \rightarrow \infty$ can only occur towards the right fix point at $H \approx 0.75 H_0$ (see Fig \eqref{add}).  Moreover, the dynamics of the case $\bar{\beta}=0$ is same as that of Case III. Hence we conclude that the late-time behavior of the considered non-linear model $f(Q)=-Q+\beta Q^2$ with $\beta \leq 0$ is similar to the linear case, whereas for the case $\beta > 0$ results are quite different.  

\section*{Acknowledgments} \label{sec8}
RS acknowledges UGC, New Delhi, India for providing Senior Research Fellowship with (UGC-Ref. No.: 191620096030). DSR acknowledges UGC, New Delhi, India for providing Junior Research Fellowship with (NTA-UGC-Ref.No.: 211610106591). SM acknowledges DST, Govt. of India, New Delhi, for providing Senior Research Fellowship (File No. DST/INSPIRE Fellowship/2018/IF18D676). PKS  acknowledges the Science and Engineering Research Board, Department of Science and Technology, Government of India for financial support to carry out the Research project No.: CRG/2022/001847. We are very much grateful to the honorable referee and to the editor for the illuminating suggestions that have significantly improved our work in terms of research quality, and presentation.

%%%%%%%%%%%%%%%%%%%%%%%%%%%%%%%%%%%%%%%%%%%%%%%%%%%%%%%%%%%%%%%%%%%%%%%%%%%%%%%%%
%%
%%

\end{document}